\documentclass[11pt]{elsarticle}

\usepackage{amssymb}
\usepackage{booktabs}
\usepackage{float}
\usepackage{diagbox}
\usepackage{lmodern}
\usepackage{amsthm}
\usepackage{graphics}
\usepackage{caption2}
\usepackage{psfrag}
\usepackage{showlabels}
\usepackage{color}
\usepackage{amsmath}
\usepackage{bm}
\usepackage{mathtools}
\usepackage{tensor}
\usepackage{subfigure}
\usepackage[normalem]{ulem}
\usepackage[linesnumbered,boxed,ruled,commentsnumbered]{algorithm2e}
\usepackage{graphicx}
\usepackage{epstopdf}
\usepackage{array}
\usepackage{amsfonts}
\usepackage{amssymb}
\usepackage{graphicx}
\usepackage{multirow}
\usepackage{exscale}
\usepackage{relsize}
\usepackage[normalem]{ulem}
\usepackage{color}
\usepackage{fullpage}
\usepackage{lineno}
\usepackage{multirow}

\newcommand{\ds}{\displaystyle}

\newcommand{\x}{\mathbf{x}}

\newcommand{\be}{\begin{equation}}
	\newcommand{\ee}{\end{equation}}
\newcommand{\ba}{\begin{array}}
	\newcommand{\ea}{\end{array}}

\newtheorem{remark}{Remark}

\begin{document}
	\begin{frontmatter}

		\title{A unified bond-based peridynamic model for isotropic and anisotropic materials}
		\author[ad1]{Hao Tian\corref{cor1}}
        \ead{haot@ouc.edu.cn}
        \author[ad1]{Jinlong Shao}
        \ead{shaojinlong@stu.ouc.edu.cn}        
        \author[ad1]{Chenguang Liu}
        \ead{liuchenguang@stu.ouc.edu.cn}
        \address[ad1]{School of Mathematical Sciences, Ocean University of China, Qingdao, Shandong 266100, China}
        \author[ad2]{Shuo Liu}
        \address[ad2]{Institute of Advanced Structure Technology, Beijing Institute of Technology, Beijing 100081, China}
        \ead{lf.lius@163.com}
        \author[ad3]{Xu Guo}
        \address[ad3]{Geotechnicaland StructuralResearchCenter,Shandong University,]inan,250061Shandong, China}
        \ead{guoxu@sdu.edu.cn}

		\begin{abstract}

           In this paper, we present a novel bond-based peridynamic model, termed Tensor-Involved Peridynamics (Ti-PD), which offers a unified framework for simulating both isotropic and anisotropic materials. This model enhances the conventional linear bond-based peridynamics by integrating a fourth-order tensor into the micromodulus function. The tensor components are calibrated to ensure the peridynamic equations converge to the classical continuum elasticity equations as the horizon parameter approaches zero. For isotropic materials with Poisson’s ratios of 1/4 in three dimensions and 1/3 in two dimensions, the Ti-PD model aligns exactly with traditional bond-based peridynamics. To further expand its applicability, we introduce a damage model specifically designed for isotropic materials, incorporating a novel critical stretch criterion distinct from ordinary state-based peridynamics. The effectiveness of the Ti-PD model in simulating general anisotropic materials is demonstrated through numerical experiments. Additionally, the damage model is validated via simulations of crack propagation in a two-dimensional plate, showcasing superior agreement with experimental data compared to conventional ordinary state-based peridynamics.

		\end{abstract}
		
		\begin{keyword}
		Peridynamics; Anisotropic materials; Elastic tensor; Zero-energy mode
		\end{keyword}	
		
	\end{frontmatter}
	\section{Introduction}
Classical continuum mechanics, which relies on partial differential equations, encounters significant challenges when modeling systems characterized by discontinuities. In response, peridynamics (PD) offers a robust nonlocal framework that effectively addresses these challenges. The original peridynamics model, classical bond-based peridynamics (BB-PD), was introduced by Silling in 2000 \cite{SA2000} and is limited to isotropic materials with a fixed Poisson's ratio. Subsequently, ordinary state-based peridynamics (OSB-PD) was proposed in 2007 \cite{SE2007}, overcoming the Poisson's ratio limitation but struggling with anisotropic materials. This paper also introduce non-ordinary state-based peridynamics (NOSB-PD), which can simulate anisotropic materials. Over the past decade, PD has stimulated extensive research in modeling \cite{RH2015,RX2017,WB2020,AV2023}, analysis \cite{Du2011,Du2013,Du2014,Ma2013,Dord2019,MA2024}, and practical applications, spanning composite material deformation \cite{OB2022,HD2015,BV2020,JC2019}, addressing corrosion \cite{CJ2021,ZJ2021,JS2019,JC20192}, predicting damage \cite{NO2021,LG2022,JD2021,NM2023,RW2022}, and simulating crack propagation in various materials \cite{SM2022,SP2017,LY2020,DP2019}.
 
In the context of anisotropic material constitutive modeling, while non-ordinary state-based peridynamics (NOSB-PD) is effective due to its capacity to incorporate constitutive relations, it is susceptible to zero-energy modes, which induce deformation under perturbations without altering the potential energy. Several methods have been developed to mitigate zero-energy modes in NOSB-PD. The first approach involves adding an additional term to the force state, a stabilization technique explored early by Littlewood \cite{Lit2010} and later by Breitenfeld \cite{BRE2014}. Building on this foundation, Silling, Li et al. \cite{SA2017}, and Wan et al. \cite{Wa2019} proposed more sophisticated stabilization schemes based on additional force states. However, these methods require parameter calibration, complicating their implementation in complex problems.
Another class of methods utilizes stabilized field state techniques. For instance, Wu \cite{WU2015} constructed the deformation gradient using averaged displacements to regularize deformation measures. Yaghoobi and Chorpeza \cite{YA2017} proposed a method incorporating higher-order terms through specific influence functions. Despite these advancements, these methods do not completely resolve the instability issue, underscoring the need for a reliable stabilization scheme. Recently, stabilization through bond association has been explored, linking kinematic quantities to PD bonds rather than to the PD material points. Notable examples include bond-level stabilization \cite{CHEN2018} and bond-associated stabilization\cite{VIEIRA2022}. A recent study \cite{FR2024} utilized an improved OSB-PD model to replace the NOSB-PD model, effectively circumventing the zero-energy mode problem.

The bond-based peridynamic model (BB-PD) is valued for its simplicity in modeling, ease of numerical implementation, and stability, as it is free from zero-energy modes. However, it is limited to isotropic materials with a fixed Poisson’s ratio. To extend its applicability, Gerstle introduced micropolar peridynamics (MPPD), which employs Euler-Bernoulli beams \cite{GERSTLE2007}. In this framework, bonds possess normal, tangential, and rotational stiffness, enabling adjustments to Poisson’s ratio.  Zheng \cite{Zheng2019} developed a bond-based PD model that incorporates rotation and shear effects, while Yu \cite{YU2020} explored the coupling of tension, rotation, and shear using Timoshenko beams. Applications of MPPD can be found in \cite{CHEN2022, CHEN2023}. Despite these advancements, the range of Poisson’s ratio remains constrained, typically below 1/4. Despite of MPPD, there have been other attempts to generalize BB-PD. Liu \cite{Liu2012} proposed a force compensation scheme to address variable Poisson’s ratios. Prakash \cite{Pra2015} introduced an enriched two-parameter elastic BB-PD model for plane stress conditions. Zhu \cite{ZHU2017} tackled these limitations with a novel BB-PD model that incorporates rotational effects, and Diana \cite{DIANA2019} proposed a  bond-based Peridynamic model with shear deformability for linear and non-linear problems by combining micropolar with this model. More research can be found in \cite{GUAN2024}, where a new force density function is introduced to distinguish between expansion and deformation, thus overcoming the limitation of Poisson's ratio. However, these models show limited applicability to anisotropic cases. A new approach for analyzing anisotropic materials within the BB-PD framework is presented in \cite{D2022}, proposing a BB-PD model characterized by six independent material micro-moduli. Shen \cite{Shen202303} constructed an anisotropic BB-PD model utilizing micro-beam bonds to simulate in-plane problem, and Liu \cite{LIU2024} introduced a conjugated BB-PD model for laminated composites featuring four independent engineering material constants. \cite{IJEE} provides a framework for calculating anisotropic materials, but it only calculates results for orthotropic anisotropy. The advancements made by the aforementioned models have significantly contributed to the generalization of bond-based peridynamics, expanding its applicability to more complex material behaviors, but limitations still persist in fully capturing the intricate characteristics of general anisotropic materials. In \cite{vito2023}, a bond-based model suitable for general anisotropic materials is established through a variational procedure. However, this model requires the additional calculation of the rotational angles of material points while considering the displacement of the material points, thus facing a heavy computational burden.

This paper presents a novel peridynamic model called tensor-involved peridynamics (Ti-PD), which effectively simulates both isotropic and anisotropic materials. Building on the framework of bond-based peridynamics (BB-PD), Ti-PD incorporates the elastic tensor into the micromodulus function, allowing for comprehensive simulation within the bond-based framework. This approach circumvents the instability associated with zero-energy modes, eliminating the need for additional stabilization methods. For isotropic material simulations with fixed Poisson's ratios (1/3 for 2D and 1/4 for 3D problems), Ti-PD can seamlessly revert to the conventional BB-PD model. Compared to existing state-based models, the proposed method significantly reduces computational complexity by leveraging the bond-based framework. Furthermore, this paper introduces a damage model for isotropic materials that is fully compatible with Ti-PD, applying energy equivalence from classical linear elastic mechanics. The model's accuracy is validated through a series of numerical experiments, demonstrating that, with direct meshfree discretization, Ti-PD can accurately capture the mechanical behavior of anisotropic materials. Additionally, for isotropic materials, Ti-PD provides a more accurate simulation of crack propagation than OSB-PD, with crack paths aligning more closely with physical experiments.

The organization of the remaining sections of this paper is as follows. Section 2 offers an overview of classical continuum mechanics and traditional peridynamics. In Section 3, we present the formulation of the 2D tensor-involved peridynamic model (Ti-PD), perform a convergence analysis relative to classical models, and derive the critical failure criterion for isotropic materials. Section 4 focuses on the model and analysis for the 3D Ti-PD. Finally, in Section 5, we validate the effectiveness and stability of the Ti-PD in both 2D and 3D anisotropic problems through a series of numerical experiments.

\section{Review of classical continuum mechanics and peridynamics}
This subsection provides a concise overview of the fundamental principles of peridynamics and classical continuum mechanics (CCM) for the sake of completeness.
\subsection{Review of CCM}
Within the Lagrangian framework, the motion equation governing each material point $\mathbf{x}$ in the domain $\Omega$ at time $t$ is given as follows \cite{CCM}.
\begin{equation}
  \rho\ddot{\mathbf{u}}(\mathbf{x},t)=\nabla\cdot \bm{\sigma}(\mathbf{x},t)+\mathbf{b}(\mathbf{x},t),
\end{equation}
Here, $\rho$ represents the density, $\mathbf{u}(\mathbf{x},t)$ denotes the displacement of the material point $\mathbf{x}$, and $\mathbf{b}(\mathbf{x},t)$ signifies the externally applied body force. Additionally, $\bm{\sigma}(\mathbf{x},t)$ corresponds to the Cauchy stress tensor.
For 2D model, using the Voigt notation, $\bm{\sigma}(\mathbf{x},t)$ can be obtained by follows\cite{ITSKOV20004} :
\begin{equation}\label{CCM2}
	\left[
	\begin{matrix}
		\sigma_{11}\\
		\sigma_{22}\\
		\sigma_{12}
	\end{matrix}
	\right]=\left[\begin{array}{llllll}
Q_{11} & Q_{12} & Q_{16}\\
& Q_{22} & Q_{26}\\
sym && D_{66}\\
\end{array}\right]
	\left[
	\begin{matrix}
		\epsilon_{11}\\
		\epsilon_{22}\\
            2\epsilon_{12} 
	\end{matrix}
	\right] 
\end{equation}
and for 3D model we have
\begin{equation}
    \left[\begin{array}{l}
\sigma_{11} \\
\sigma_{22} \\
\sigma_{33} \\
\sigma_{12} \\
\sigma_{23} \\
\sigma_{13}
\end{array}\right]=\left[\begin{array}{llllll}
Q_{11} &Q_{12} &Q_{13} &Q_{14} &Q_{15} &Q_{16}\\
  &Q_{22}&Q_{23}&Q_{24}&Q_{25}&Q_{26} \\
 &&Q_{33}&Q_{34}&Q_{35}&Q_{36} \\
&&&Q_{44}&Q_{45}&Q_{46}\\
&&&&Q_{55}&Q_{56}\\
sym&&&&&D_{66}
\end{array}\right]\left[\begin{array}{c}
\varepsilon_{11} \\
\varepsilon_{22} \\
\varepsilon_{33} \\
2 \varepsilon_{12} \\
2 \varepsilon_{23} \\
2 \varepsilon_{13}
\end{array}\right].
\end{equation}

\subsection{Fundamentals of peridynamics}

Peridynamics is founded on nonlocal concepts, asserting that a material point interacts within a finite region $ \mathbf{\mathcal{B}_{\delta}(\mathbf{x})}=\left\{\mathbf{x}^{\prime} \mid\left\|\mathbf{x}^{\prime}-\mathbf{x}\right\|<\delta\right\}$, and reformulates classical continuum solid mechanics using the integral form instead of partial differential equations.
The motion equation for PD \cite{SA2000}  can be defined as follows
    \begin{equation}\label{pd:e1}
    \vspace{0.01in}
    \rho \ddot{\mathbf{u}}(\mathbf{x}, t)= \ds \int_{\mathbf{\mathcal{B}_{\delta}(\mathbf{x})}} \mathbf{f}(\bm{\eta},\bm{\xi}) \mathrm{d}V_{\mathbf{\mathbf{x}^{\prime}}}+\mathbf{b}(\mathbf{x},t), 
    \end{equation}
where $\rho$ represents the mass density, $\mathbf{b}(\mathbf{x},t)$ denotes the body force density field, $\bm{\xi}=\mathbf{x}^{'}-\mathbf{x}$, $\bm{\eta}=\mathbf{u}(\mathbf{x}^{'},t)-\mathbf{u}(\mathbf{x},t)$. 

The function $\mathbf{f}(\bm{\eta},\bm{\xi})$ represents a pairwise force function utilized to calculate the force vector. Importantly, it satisfies the antisymmetry condition $\mathbf{f}(\bm{\eta},\bm{\xi}) = -\mathbf{f}(-\bm{\eta},-\bm{\xi})$. Depending on the specific form of $\mathbf{f}$, the primary peridynamic formulations can be classified into three categories: bond-based peridynamics (BB-PD), ordinary state-based peridynamics (OSB-PD), and non-ordinary state-based peridynamics (NOSB-PD).

For  BB-PD, force density function can be expressed as
\begin{equation}\label{BB-PDlin}
\mathbf{f}(\bm{\eta},\bm{\xi})=\alpha \frac{\mathbf{\bm{\xi}}\otimes\mathbf{\bm{\xi}}}{\|\mathbf{\bm{\xi}}\|^{3}}\bm{\eta}, \quad \alpha= \displaystyle{\begin{cases}\dfrac{9E}{\pi\delta^{3}h},& \text {2D}, \vspace{0.5em}\\\dfrac{12E}{\pi\delta^{4}},&\text {3D}. \\ \end{cases}}
\end{equation}

For OSB-PD, we have
	\begin{equation}
	\mathbf{f}(\bm{\eta},\bm{\xi})=\beta \omega(\|\bm{\xi}\|)\left(\theta^l(\mathbf{x}^{\prime})-\theta^l(\mathbf{x})\right) \boldsymbol{\xi}+2\gamma \omega(\|\bm{\xi}\|)  \frac{\boldsymbol{\xi} \otimes \boldsymbol{\xi}}{\|\boldsymbol{\xi}\|^2} \bm{\eta},
	\end{equation}
with
\begin{equation}
    \beta=\frac{3K-5G}{m},\quad \gamma=-\frac{30G}{m},\quad 
 \theta^l(\mathbf{x})=\frac{3}{m} \int_{\mathbf{\mathcal{B}_{\delta}(\mathbf{x})}} \omega(\|\boldsymbol{\xi}\|) \boldsymbol{\xi} \cdot \boldsymbol{\eta} \mathrm{d} V_{\boldsymbol{x}^{\prime}}.
\end{equation}
where $G$ represents the shear modulus, $K$ denotes the bulk modulus, and
	\begin{equation}
	m=\int_{\mathbf{\mathcal{B}_{\delta}(\mathbf{x})}} \omega(\|\bm{\xi}\|)\|\boldsymbol{\xi}\|^2 \mathrm{~d} V_{\boldsymbol{x}^{\prime}},\quad \omega(\|\bm{\xi}\|)=\dfrac{1}{\|\bm{\xi}\|}.
	\end{equation}

For NOSB-PD, we have
\begin{equation}\label{NOSB}
   \mathbf{f}(\bm{\eta},\bm{\xi})=\mathbf{T}[\mathbf{x}, t]\langle\bm{\xi}\rangle-\mathbf{T}\left[\mathbf{x}^{\prime}, t\right]\langle-\boldsymbol{\xi}\rangle,
\end{equation}
with
\begin{equation}
\mathbf{T}[\mathbf{x}, t]\langle\bm{\xi}\rangle=\omega(\|\bm{\xi}\|)(\mathcal{C}:\bm{\epsilon})\cdot\mathbf{K}^{-1}\cdot\bm{\xi}, \quad \mathbf{K}=\int_{\mathbf{\mathcal{B}_{\delta}(\mathbf{x})}} \omega(\|\bm{\xi}\|)(\bm{\xi} \otimes \bm{\xi}) \mathrm{d} V_{\mathbf{x}^{\prime}},
\end{equation}

where $\mathcal{C}$ is the isotropic elastic moduli matrix and
\begin{equation}
  \bm{\epsilon}=\dfrac{\nabla^N\mathbf{u}+(\nabla^N\mathbf{u})^T}{2},\quad \nabla^N\mathbf{u}=\int_{\mathcal{B}_{\delta}(\x)} \underline{\omega}(\|\bm{\xi}\|) \bm{\eta} \otimes \bm{\xi}\mathrm{d} V_{\mathbf{x}^{\prime}} \cdot \mathbf{K}^{-1} 
\end{equation}

Compared to the BB-PD model, the OSB-PD and NOSB-PD models offer a broader range of applications. However, the construction of double integrals introduces significant computational complexity, which can be prohibitive for practical calculations. Furthermore, the presence of zero-energy modes poses a critical challenge to the effective application of NOSB-PD.

\section{2D Tensor-involved peridynamics}
Existing peridynamic (PD) models encounter several challenges. The BB-PD model is limited to isotropic materials with a fixed Poisson's ratio, while both OSB-PD and NOSB-PD entail significant computational costs. Furthermore, when dealing with anisotropic materials, the need for correction methods to address the zero-energy mode in the NOSB-PD model further increases this computational burden. To address these issues, we propose a 2D Ti-PD model in this section, designed to simulate both isotropic and anisotropic materials within a single framework. This model specifically aims to overcome the limitations of existing PD models in simulating anisotropic materials. Additionally, we introduce a fracture model associated with Ti-PD, based on energy equivalence.

\subsection{Formulations of 2D Ti-PD}

In this subsection, we construct a novel peridynamic model by incorporating elastic tensor information into the micromodulus function of the traditional BB-PD. The resulting equation of motion is given by:
\begin{equation}
		\rho \ddot{\mathbf{u}}(\mathbf{x}, t)= \mathcal{L}\mathbf{u}(\mathbf{x}, t)+\mathbf{b}(\mathbf{x},t)
\end{equation}
where the $\mathcal{L}\mathbf{u}(\mathbf{x}, t)$ is the operate for the Ti-PD and can be expressed as
	\begin{equation}\label{newL3d}
			\mathcal{L}\mathbf{u}(\mathbf{x}, t)= \ds \int_{\mathbf{\mathcal{B}_{\delta}(\mathbf{x})}} \mathbf{f}(\bm{\eta},\bm{\xi}) dV_{\mathbf{\mathbf{x}^{\prime}}}
	\end{equation}
with
\begin{equation}\label{g2}
 \mathbf{f}(\bm{\eta},\bm{\xi}) =  \frac{1}{\pi \delta^3 h}\dfrac{\mathcal{D}:(\bm{\xi}\otimes\bm{\xi})}{\|\mathbf{\bm{\xi}}\|^{3}}\bm{\eta},
\end{equation}
The matrix of introduced fourth-order tensor $\mathcal{D}=D^{i\hspace{0.12em} k}_{.j.l}\mathbf{g}_i\otimes\mathbf{g}_j\otimes\mathbf{g}_k\otimes\mathbf{g}_l$ can be expressed as the following matrix:
\begin{align}\label{new2d}
\mathbf{D}&=\left[\begin{array}{lllllllll}
D^{1\hspace{0.12em}1}_{.1.1} & D^{1\hspace{0.12em}1}_{.1.2} & D^{1\hspace{0.12em}2}_{.1.1} & D^{1\hspace{0.12em}2}_{.1.2}\\
D^{1\hspace{0.12em}1}_{.2.1} & D^{1\hspace{0.12em}1}_{.2.2}&D^{1\hspace{0.12em}2}_{.2.1} & D^{1\hspace{0.12em}2}_{.2.2}\\
D^{2\hspace{0.12em}1}_{.1.1} & D^{2\hspace{0.12em}1}_{.1.2} & 
D^{2\hspace{0.12em}2}_{.1.1} & D^{2\hspace{0.12em}2}_{.1.2}\\
D^{2\hspace{0.12em}1}_{.2.1} & D^{2\hspace{0.12em}1}_{.2.2} & D^{2\hspace{0.12em}2}_{.2.1} & D^{2\hspace{0.12em}2}_{.2.2}
\end{array}\right].
\end{align}
assuming that two symmetry
 $D$ satisfies $D_{.il}^{jk}=D_{.ik}^{jl}$, $D_{.il}^{jk}=D_{.jk}^{il}$, $D$ can be rewritten in the following form:

\begin{align}
\mathbf{C}&=
\mathcal{D}:(\mathbf{\bm{\xi}}\otimes\mathbf{\bm{\xi}})
=\left[\begin{matrix}
    C_{11}&C_{12}\\
    C_{21}&C_{22}
\end{matrix}\right]\\
&=\left[\begin{array}{llllll}
    D^{1\hspace{0.12em}1}_{.1.1}\xi_1^2+ 2D^{1\hspace{0.12em}1}_{.1.2}\xi_1\xi_2+ D^{1\hspace{0.12em}2}_{.1.2}\xi_2^2&
    D^{1\hspace{0.12em}1}_{.2.1}\xi_1^2+ 2D^{1\hspace{0.12em}1}_{.2.2}\xi_1\xi_2 + D^{1\hspace{0.12em}2}_{.2.2}\xi_2^2\\
    D^{1\hspace{0.12em}1}_{.2.1}\xi_1^2+ 2D^{1\hspace{0.12em}1}_{.2.2}\xi_1\xi_2 + D^{1\hspace{0.12em}2}_{.2.2}\xi_2^2&
D^{2\hspace{0.12em}1}_{.2.1}\xi_1^2+ 2D^{2\hspace{0.12em}1}_{.2.2} \xi_1\xi_2+D^{2\hspace{0.12em}2}_{.2.2}\xi_2^2\end{array}\right].
\end{align}

The  process for solving $\mathcal{D}$ can be obtained as follows:

When $\delta\rightarrow0$, the peridynamic equation converge to Navier’s equation, thus we can obtain the the linear constitutive equation \cite{IJEE}.
 Using the Taylor expansion, $\mathbf{u}(\mathbf{x},t)=(u(\mathbf{x},t) , v(\mathbf{x},t))^T$ can be rewritten as :
 \begin{equation}\label{taly}
   \bm{\eta}=\mathbf{u}(\mathbf{x`}^{\prime},t)-\mathbf{u}(\mathbf{x},t)=\sum_{n=1}^\infty\frac1{n!}\Bigg(\xi_1\frac\partial{\partial x_1}+\xi_2\frac\partial{\partial x_2}\Bigg)^n\mathbf{u}(\mathbf{x},t)
\end{equation}

Let $P(\mathbf{x},t)=\mathcal{L}\mathbf{u}(\mathbf{x}, t)$,  using \eqref{taly}, we can obtain
\begin{equation}
P(\mathbf{x}, t)=\sum_{n=1}^\infty\frac1{n!}P_n(\mathbf{x}, t)\end{equation}
where
\begin{align}
P_n(\mathbf{x}, t)& =\frac{1}{\pi \delta^3 h}\ds \int_{\mathbf{\mathcal{B}_{\delta}(\mathbf{x})}}\dfrac{\mathbf{C}(\boldsymbol{\xi}\bullet\nabla)^n\mathbf{\bm{u}}}{\|\mathbf{\bm{\xi}}\|^{3}}dV_{\mathbf{\mathbf{x}^{\prime}}} \\
&=\frac{1}{\pi \delta^3 h}\ds \int_{\mathbf{\mathcal{B}_{\delta}(\mathbf{x})}}\dfrac{(\boldsymbol{\xi}\bullet\nabla)^n\begin{pmatrix}C_{11}u+
    C_{12}v\\ C_{21}u+
C_{22}v\end{pmatrix}}{\|\mathbf{\bm{\xi}}\|^{3}}dV_{\mathbf{\mathbf{x}^{\prime}}}
\end{align}
  It can be shown that
\begin{equation}P_{2m-1}(\mathbf{x}, t)=\frac{1}{\pi \delta^3 h}\ds \int_{\mathbf{\mathcal{B}_{\delta}(\mathbf{x})}}\dfrac{\mathbf{C}(\boldsymbol{\xi}\bullet\nabla)^{2m-1}\mathbf{\bm{u}}}{\|\mathbf{\bm{\xi}}\|^{3}}dV_{\mathbf{\mathbf{x}^{\prime}}}=0\quad(m\geq1)\end{equation}
$\mathbf{C}(\xi\bullet\nabla)^{2m-1}\bm{u}$ is an odd function of at least one variable of $\boldsymbol{\xi}=(\xi_1,\xi_2)$, and the integration domain is symmetric with respect to the origin. Thus, using (30) in (38), we can obtain
\begin{equation}P(\mathbf{x}, t)=\sum_{m=1}^\infty\frac1{(2m)!}P_{2m}(\mathbf{x}, t)\end{equation}
Here $\nabla$ is the del operator. Then the following equation can be proved
\begin{equation}\lim_{\delta\to0}P_{2m}(\mathbf{x}, t)=0\quad\mathrm{if~}m\geq2\end{equation}
Let  $u_{i,j}=\dfrac{\partial u^2}{\partial i \partial j}$,$v_{i,j}=\dfrac{\partial v^2}{\partial i \partial j}$, and  
since
\begin{equation}\label{2Dint}
\int_{\mathbf{\mathcal{B}_{\delta}(\mathbf{x})}}\frac{\xi_1^a\xi_2^{3-a}}{\|\mathbf{\bm{\xi}}\|^{3}} dV_{\mathbf{x^{\prime}}} = 0, \quad a = 0, 1, 2, 3;
\end{equation}
\begin{equation}
\int_{\mathcal{B}_\delta(\mathbf{x})} \frac{\xi_1^a\xi_2^{4-a}}{\left\|\bm{\xi}\right\|^3} d V_{\mathbf{x}^{\prime}}= \begin{cases}\frac{\pi \delta^3}{4}, & a=0,4 ; \\ 0, & a=1,3 ; \\ \frac{\pi \delta^3}{12}, & a=2 ;\end{cases},
\end{equation}
when $\delta \to 0$, by substituting (31) into (30) , we obtain 
 {\smaller\begin{align}\label{tipdccm}
 &\lim_{\delta \to 0}P(\mathbf{x}, t)=\lim_{\delta \to 0}\frac1{2}P_2(\mathbf{x}, t)=\lim_{\delta \to 0}\frac{1}{2\pi \delta^3 h}\ds \int_{\mathbf{\mathcal{B}_{\delta}(\mathbf{x})}}\dfrac{\mathbf{C}(\boldsymbol{\xi}\bullet\nabla)^{2}\mathbf{\bm{u}}}{\|\mathbf{\bm{\xi}}\|^{3}}dV_{\mathbf{\mathbf{x}^{\prime}}}\\&=\lim_{\delta \to 0}\frac{1}{2\pi \delta^3 h}\ds \int_{\mathbf{\mathcal{B}_{\delta}(\mathbf{x})}}\dfrac{1}{\|\mathbf{\bm{\xi}}\|^{3}}\left[\begin{array}{llllll}
    D^{1\hspace{0.12em}1}_{.1.1}\xi_1^2+ 2D^{1\hspace{0.12em}1}_{.1.2}\xi_1\xi_2+ D^{1\hspace{0.12em}2}_{.1.2}\xi_2^2&
    D^{1\hspace{0.12em}1}_{.2.1}\xi_1^2+ 2D^{1\hspace{0.12em}1}_{.2.2}\xi_1\xi_2 + D^{1\hspace{0.12em}2}_{.2.2}\xi_2^2\\
    D^{1\hspace{0.12em}1}_{.2.1}\xi_1^2+ 2D^{1\hspace{0.12em}1}_{.2.2}\xi_1\xi_2 + D^{1\hspace{0.12em}2}_{.2.2}\xi_2^2&
D^{2\hspace{0.12em}1}_{.2.1}\xi_1^2+ 2D^{2\hspace{0.12em}1}_{.2.2} \xi_1\xi_2+D^{2\hspace{0.12em}2}_{.2.2}\xi_2^2\end{array}\right]\left[\begin{matrix}
     \xi_1^2u_{xx}+2\xi_1\xi_2u_{xy}+\xi_2^2u_{yy}\\ \xi_1^2v_{xx}+2\xi_1\xi_2v_{xy}+\xi_2^2v_{yy}
 \end{matrix}\right]dV_{\mathbf{\mathbf{x}^{\prime}}}\\&=\left[\begin{matrix}
     \dfrac{3D_{.1.1}^{11}+D_{.1.2}^{12}}{24}u_{xx}+\dfrac{D_{.1.2}^{11}}{6}u_{xy}+  \dfrac{D_{.1.1}^{11}+3D_{.1.2}^{12}}{24}u_{yy}+  \dfrac{3D_{.2.1}^{11}+D_{.2.2}^{12}}{24}v_{xx}+  \dfrac{D_{.2.2}^{11}}{6}v_{xy}+  \dfrac{D_{.2.1}^{11}+3D_{.2.2}^{12}}{24}v_{yy}\\     \dfrac{3D_{.2.1}^{11}+D_{.2.2}^{12}}{24}u_{xx}+\dfrac{D_{.2.2}^{11}}{6}u_{xy}+  \dfrac{D_{.2.1}^{11}+3D_{.2.2}^{12}}{24}u_{yy}+  \dfrac{3D_{.2.1}^{21}+D_{.2.2}^{22}}{24}v_{xx}+  \dfrac{D_{.2.2}^{21}}{6}v_{xy}+  \dfrac{D_{.2.1}^{21}+3D_{.2.2}^{22}}{24}v_{yy}
 \end{matrix}\right]\end{align}}


Substitute  \eqref{taly} and \eqref{2Dint} into (1), we have the following equations  according to the first component in \eqref{tipdccm}:
\begin{align}\label{eqx}
    \frac{D_{.1.1}^{11}}{8}+\frac{D_{.1.2}^{12}}{2
    4}=Q_{11}\notag\\
    \frac{D_{.1.2}^{11}}{6}=2Q_{16}\notag\\
    \frac{D_{.1.1}^{11}}{24}+\frac{D_{.1.2}^{12}}{8}=Q_{66}\notag\\
    \frac{D_{.2.1}^{11}}{8}+\frac{D_{.2.2}^{12}}{2
    4}=Q_{16}\\
    \frac{D_{.2.2}^{11}}{6}=Q_{12}+Q_{66}\notag\\
    \frac{D_{.2.1}^{11}}{24}+\frac{D_{.2.2}^{12}}{8}=Q_{26}\notag
\end{align}

and we can obtain following equations by the second component:
\begin{align}\label{eqy}
\frac{D_{.2.1}^{11}}{8}+\frac{D_{.2.2}^{12}}{24}=Q_{16}\notag\\
\frac{D_{.2.2}^{11}}{6}=Q_{12}+Q_{66}\notag\\
\frac{D_{.2.1}^{11}}{24}+\frac{D_{.2.2}^{12}}{8}=Q_{26}\notag\\
    \frac{D_{.2.1}^{21}}{8}+\frac{D_{.2.2}^{22}}{2
    4}=Q_{66}\\
    \frac{D_{.2.2}^{21}}{6}=2Q_{26}\notag\\
    \frac{D_{.2.1}^{21}}{24}+\frac{D_{.2.2}^{22}}{8}=Q_{22}\notag
\end{align}

 Since the first three equations in \eqref{eqy} are equal to the last three equations in \eqref{eqx}, those 12 equations can be simplified to 9 equations. By solving these 9 equations, the specific form of $D$ can be obtained:
\begin{align}
    D=\left.\left[\begin{array}{llll}9Q_{11}-3Q_{66}&12Q_{16}&12Q_{16}&9Q_{66}-3Q_{11}\\9Q_{16}-3Q_{26}&6(Q_{12}+Q_{66})&6(Q_{12}+Q_{66})&9Q_{26}-3Q_{16}\\9Q_{16}-3Q_{26}&6(Q_{12}+Q_{66})&6(Q_{12}+Q_{66})&9Q_{26}-3Q_{16}\\9Q_{66}-3Q_{22}&12Q_{26}&12Q_{26}&9Q_{22}-3Q_{66}\end{array}\right.\right].
\end{align}

To numerically implement the proposed Ti-PD model, we discretize it using the meshfree approach described in \cite{SA2005}. In this method, the domain is represented by a collection of nodes, where each node $\mathbf{x_p}$ is associated with a known volume $V_p$. Let $\mathbf{x_p}$ and $\mathbf{x_q}$ denote two distinct material points, and $\mathbf{u_p}$ and $\mathbf{u_q}$ represent their respective displacements. The governing equation (43) can then be expressed in its discrete form as follows:
\begin{equation}\label{meshfree} \rho \ddot{\mathbf{u}}(\mathbf{x}_p, t) = \sum \mathbf{A}_{p,q}(\mathbf{u_q} - \mathbf{u_p})V_q, \end{equation}
where the stiffness matrix $\mathbf{A}$ is shown in Appendix B.

\begin{remark}
Unlike the elastic tensor in classical continuum mechanics (CCM), the tensor introduced in Ti-PD satisfies only two symmetries rather than the three symmetries, which is $D^{i\hspace{0.12em}k}_{.j.l} = D^{j\hspace{0.12em}k}_{.i.l}$ and $D^{i\hspace{0.12em}k}_{.j.l} = D^{i\hspace{0.12em}l}_{.j.k}$. In fact, if the introduced tensor satisfies
$D^{i\hspace{0.12em}k}_{.j.l} = D^{k\hspace{0.12em}i}_{.l.j}$, and when equations \eqref{eqx} and \eqref{eqy} have a unique solution, the elements in the tensor must satisfy the relations $Q_{26}+Q_{16}=0$, $Q_{11}-Q_{22}=0$. Thus, the model constructed by enforcing three types of symmetry cannot accurately simulate general anisotropic materials.

\end{remark}


 \begin{remark}
    The Ti-PD model is versatile enough to be applied to general anisotropic materials and can also handle isotropic materials effectively. For isotropic cases, the fourth-order elastic tensor in equation (\ref{new2d}) can be expressed as:
    \begin{equation}\label{2DOSB-PDD}
    \mathbf{D} = \frac{E}{1 - \nu^2}
        \left[
        \begin{matrix}
            \dfrac{15+3\nu}{2} & 0 & 0 & \dfrac{3-9\nu}{2} \\
            0 & 3(1 + \nu) & 3(1 + \nu) & 0 \\
            0 & 3(1 + \nu) & 3(1 + \nu) & 0 \\
            \dfrac{3-9\nu}{2} & 0 & 0 & \dfrac{15+3\nu}{2}
        \end{matrix}
        \right]
    \end{equation}
    Specifically, when Poisson's ratio is taken as \(\nu = 1/3\), the tensor simplifies to:
    \begin{equation}\label{2DBB-PDD}
    \mathbf{D} = 9E
        \left[
        \begin{matrix}
            1 & 0 & 0 & 0 \\
            0 & \dfrac{1}{2} & \dfrac{1}{2} & 0 \\
            0 & \dfrac{1}{2} & \dfrac{1}{2} & 0 \\
            0 & 0 & 0 & 1
        \end{matrix}
        \right]
    \end{equation}
    Substituting equation \eqref{2DBB-PDD} into equation \eqref{g2}, we obtain:
    \begin{equation}
    \mathbf{f}(\bm{\eta}, \bm{\xi}) = \frac{9E}{\pi \delta^3 h} \frac{\mathbf{\bm{\xi}} \otimes \mathbf{\bm{\xi}}}{\|\mathbf{\bm{\xi}}\|^{3}} \bm{\eta}
    \end{equation}
    This corresponds exactly to the traditional bond-based peridynamics (BB-PD) formulation \eqref{BB-PDlin}, reinforcing the notion that the proposed model serves as an extension of the BB-PD framework. The Ti-PD model can revert to the classical BB-PD approach when applied to isotropic materials with \(\nu = 1/3\).
\end{remark}

\subsection{The critical failure criterion of Ti-PD for 2D isotropic materials}
In this section, we explore the concept of failure in the Ti-PD framework for isotropic materials. The most straightforward approach to model failure is to permit bonds to break when they are extended beyond a specified threshold. For illustrative purposes, we introduce a history-dependent scalar-valued function akin to that used in traditional peridynamics:
\begin{equation}
    \mathcal{L}\mathbf{u}(\mathbf{x}, t) = \int_{\mathcal{B}_{\delta}(\mathbf{x})} \mu(\mathbf{x}, \mathbf{x}^{\prime}, t) \mathbf{f}(\bm{\eta}, \bm{\xi}) \, dV_{\mathbf{x}^{\prime}}, \quad \mu(\mathbf{x}, \mathbf{x}^{\prime}, t) = 
    \begin{cases} 
    1 & \text{if } s < s_0 \\ 
    0 & \text{if } s \ge s_0 
    \end{cases}
\end{equation}
Here, \(s_0\) is the critical stretch for bond failure, and \(s\) is defined as the bond stretch:
\begin{equation}
    s = \frac{\|\bm{\xi} + \bm{\eta}\| - \|\bm{\xi}\|}{\|\bm{\xi}\|}.
\end{equation}
We use  \(\mu(\mathbf{x}, \mathbf{x}^{\prime}, t)\) to clarify the concept of local damage:

	 \begin{equation}
	\varphi(\mathbf{x},t)=1-\frac{\ds\int_{\mathbf{\mathcal{B}_{\delta}(\mathbf{x})}}\mu(\mathbf{x},\mathbf{x}^{\prime},t)dV_{\mathbf{\mathbf{x}}}}{\ds\int_{\mathbf{\mathcal{B}_{\delta}(\mathbf{x})}}dV_{\mathbf{\mathbf{x}}}}	
\end{equation}
where $0\leq\varphi(\mathbf{x},t)\leq1$, a value of 0 indicates an undamaged state, while a value of 1 signifies complete separation of the material point $\mathbf{x}$ from all points within its influence horizon.

\begin{figure}[ht]
		\centering            
		\includegraphics[scale=0.5]{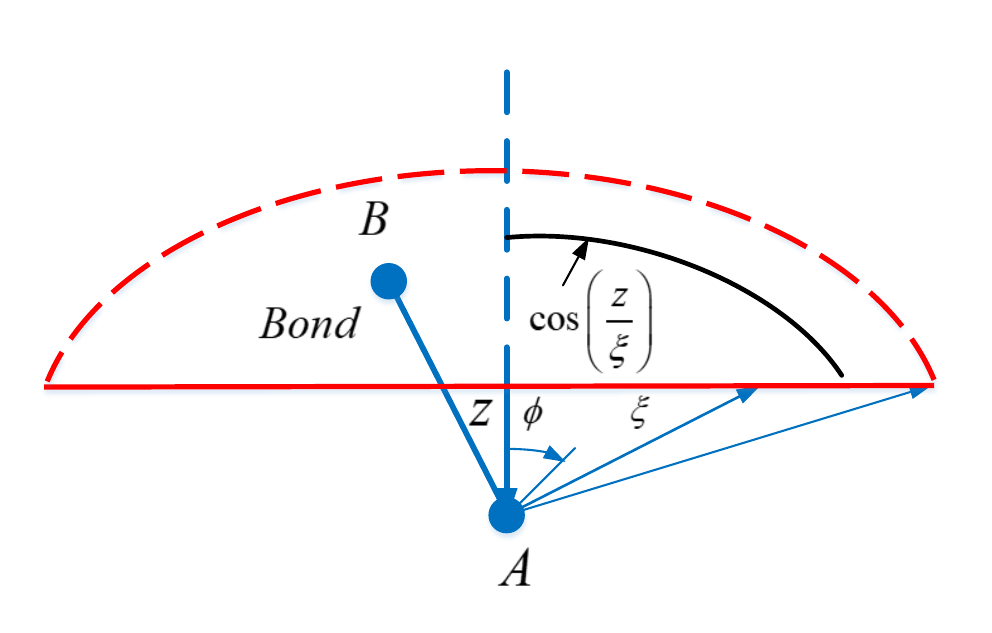}  
		\caption{Evaluation of fracture energy $G_0$  in 3D model for each material point $\mathbf{A}$. Here red real line is the fracture surface.  $\mathbf{A}$ and  $\mathbf{B}$ are material points and  $0\leq z \leq \delta$. The energy required to break the bond between material point $\mathbf{A}$ and material point $\mathbf{B}$ in a circular coordinate system can be derived through equation \eqref{2DG0}.} 
		\label{Fig:s0}
	\end{figure}

Consider an isotropic homogeneous body, 
where $s$ is constant for all $\xi$, and $\bm{\eta}=s \bm{\xi}$. Let $\xi=|\bm{\xi}|$ and $\eta=|\boldsymbol{\eta}|$, 
we can obtain $\eta=s \xi$.  The failure constant $s_0$ is related to the work required to break a single bond. Here  $G_0$ is introduced to denote the energy required to fracture all bonds per unit area. For the 2D model, we have \cite{HA2011}
\begin{equation}\label{2DG0}
 	G_0= \ds \int_{0}^{\delta}\int_{z}^{\delta}\int_{0}^{cos^{-1} z/\xi} \mathbf{f}(\bm{\eta},\bm{\xi})\cdot\bm{\eta}\xi d\phi d\xi dz =\frac{3Es_0^2\delta(5+\nu)}{4\pi(1-\nu^2)}
 \end{equation}
 An explanation of this computation are shown in Fig. \ref{Fig:s0}. 

 Thus, solving (\refeq{2DG0}) for the critical bond stretch $s_0$ lead to

 \begin{equation}\label{2DOSB-PDs0}
 	s_0=\sqrt{\frac{8\pi G_0(1-v^2)}{3E\delta(5+v)}}
 \end{equation}
 More generally, when Poisson's ratio is $1/3$, the expression of $s_0$ can be obtained by (\refeq{2DOSB-PDs0}) as:
  \begin{equation}\label{2DBB-PDs0}
 	s_0=\sqrt{\frac{4\pi G_0}{9E\delta}}
 \end{equation}
which means that the new critical bond stretch is equal to $s_0$ in the two-dimensional BB-PD model when simulate isotropic materials. This further validates the consistency between our proposed model and BB-PD when dealing with isotropic materials.
 \begin{remark}
     For OSB-PD, the the expression of $s_0$ can be derived as follows:
\begin{equation}
s_0=\sqrt{\frac{G_0}{\left(\frac{6}{\pi} \mu+\frac{16}{9 \pi^2}(\kappa-2G)\right) \delta}} =\sqrt{\frac{9\pi^2(1+\nu)(1-\nu)G}{(8(3\nu-1)+27\pi(1-\nu))E\delta}}
\end{equation}
with $G$ is the shear modulus, $\kappa$ is the bulk modulus.

  It is clear that when the Poisson's ratio is not equal to \( 1/3 \), the critical bond stretch \( s_0 \) obtained from the Ti-PD model differs from that derived from OSB-PD. Nonetheless, using the Ti-PD model with \( s_0 \) from OSB-PD still produces more accurate results than OSB-PD alone. Moreover, employing the newly derived \( s_0 \) further improves accuracy. Additional details can be found in Section 5.3.

 \end{remark}
  
\begin{remark}
The aforementioned concept of failure is applicable only to isotropic materials. The damage criteria for anisotropic materials within the Ti-PD framework require further investigation. Currently, when simulating anisotropic materials, Ti-PD can provide stable and accurate displacements, allowing us to employ the same failure criteria as those proposed for NOSB-PD, as demonstrated in \cite{Do2024}.
\end{remark}
\section{3D Tensor-involved peridynamics}

In this section, we discuss the motion equation and damage model of 3D Ti-PD.
\subsection{Motion equation}
we consider the Ti-PD for 3D anisotropy material, we first provide the motion equation form of a three-dimensional problem in classical continuum mechanics. Similiar to 2D model, the force density vector for 3D model can be written as:
 \begin{equation}\label{g23d}
		\mathbf{f}(\bm{\eta},\bm{\xi})=\frac{1}{\pi \delta^4}\dfrac{\mathbf{\mathcal{D}}:(\bm{\xi}\otimes\bm{\xi})}{\|\mathbf{\bm{\xi}}\|^{3}} \bm{\eta}
\end{equation}
Here $D$ is a fouth-tensor and the matrix representation can be expressed as:
\begin{equation}
\mathbf{D}=\left[\begin{array}{lllllllll}
D^{1\hspace{0.12em}1}_{.1.1} & D^{1 \hspace{0.12em}2}_{.1.1} & D^{1\hspace{0.12em}3}_{.1.1} & D^{1\hspace{0.12em}1}_{.1.2} & D^{1\hspace{0.12em}2}_{.1.2} & D^{1\hspace{0.12em}3}_{.1.2} & D^{1\hspace{0.12em}1}_{.1.3} & D^{1\hspace{0.12em}2}_{.1.3} & D^{1\hspace{0.12em}3}_{.1.3}\\
D^{1\hspace{0.12em}1}_{.2.1} & D^{1\hspace{0.12em}2}_{.2.1} & D^{1\hspace{0.12em}3}_{.2.3}&D^{1\hspace{0.12em}1}_{.2.2} & D^{1\hspace{0.12em}2}_{.2.2} & D^{1\hspace{0.12em}3}_{.2.2} &D^{1\hspace{0.12em}1}_{.2.3} & D^{1\hspace{0.12em}2}_{.2.3} & D^{1\hspace{0.12em}3}_{.2.3}\\
D^{1\hspace{0.12em}1}_{.3.1} & D^{1\hspace{0.12em}2}_{.3.1} & D^{1\hspace{0.12em}3}_{.3.1}&D^{1\hspace{0.12em}1}_{.3.2} & D^{1\hspace{0.12em}2}_{.3.2} & D^{1\hspace{0.12em}3}_{.3.2}&D^{1\hspace{0.12em}1}_{.3.3} & D^{1\hspace{0.12em}2}_{.3.3} & D^{1\hspace{0.12em}3}_{.3.3}\\
D^{2\hspace{0.12em}1}_{.1.1} & D^{2\hspace{0.12em}2}_{.1.1} & D^{2\hspace{0.12em}3}_{.1.1}&D^{2\hspace{0.12em}1}_{.1.2} & D^{2\hspace{0.12em}2}_{.1.2} & D^{2\hspace{0.12em}3}_{.1.2}&D^{2\hspace{0.12em}1}_{.1.3} & D^{2\hspace{0.12em}2}_{.1.3} & D^{2\hspace{0.12em}3}_{.1.3}\\
D^{2\hspace{0.12em}1}_{.2.1} & D^{2\hspace{0.12em}2}_{.2.1} & D^{2\hspace{0.12em}3}_{.2.3}&D^{2\hspace{0.12em}1}_{.2.2} & D^{2\hspace{0.12em}2}_{.2.2} & D^{2\hspace{0.12em}3}_{.2.2}& D^{2\hspace{0.12em}1}_{.2.3} & D^{2\hspace{0.12em}2}_{.2.3} & D^{2\hspace{0.12em}3}_{.2.3}\\
D^{2\hspace{0.12em}1}_{.3.1} & D^{2\hspace{0.12em}2}_{.3.1} & D^{2\hspace{0.12em}3}_{.3.1}&D^{2\hspace{0.12em}1}_{.3.2} & D^{2\hspace{0.12em}2}_{.3.2} & D^{2\hspace{0.12em}3}_{.3.2}&D^{2\hspace{0.12em}1}_{.3.3} & D^{2\hspace{0.12em}2}_{.3.3} & D^{2\hspace{0.12em}3}_{.3.3}\\
D^{3\hspace{0.12em}1}_{.1.1} & D^{3\hspace{0.12em}2}_{.1.1} & D^{3\hspace{0.12em}3}_{.1.1}& D^{3\hspace{0.12em}1}_{.1.2} & D^{3\hspace{0.12em}2}_{.1.2} & D^{3\hspace{0.12em}3}_{.1.2}&D^{3\hspace{0.12em}1}_{.1.3} & D^{3\hspace{0.12em}2}_{.1.3} & D^{3\hspace{0.12em}3}_{.1.3}\\
D^{3\hspace{0.12em}1}_{.2.1} & D^{3\hspace{0.12em}2}_{.2.1} & D^{3\hspace{0.12em}3}_{.2.3}&D^{3\hspace{0.12em}1}_{.2.2} & D^{3\hspace{0.12em}2}_{.2.2} & D^{3\hspace{0.12em}3}_{.2.2}&D^{3\hspace{0.12em}1}_{.2.3} & D^{3\hspace{0.12em}2}_{.2.3} & D^{3\hspace{0.12em}3}_{.2.3}\\
D^{3\hspace{0.12em}1}_{.3.1} & D^{3\hspace{0.12em}2}_{.3.1} & D^{3\hspace{0.12em}3}_{.3.1}&D^{3\hspace{0.12em}1}_{.3.2} & D^{3\hspace{0.12em}2}_{.3.2} & D^{3\hspace{0.12em}3}_{.3.2}&D^{3\hspace{0.12em}1}_{.3.3} & D^{3\hspace{0.12em}2}_{.3.3} & D^{3\hspace{0.12em}3}_{.3.3}
\end{array}\right]
\end{equation}
assuming that $D$ satisfies $D_{.il}^{jk}=D_{.ik}^{jl},D_{.il}^{jk}=D_{.jk}^{il}$, we have
\begin{align}
\mathbf{C}&=
\mathbf{D}:(\mathbf{\bm{\xi}}\otimes\mathbf{\bm{\xi}})
=\left[\begin{matrix}
    C_{11}&C_{12}&C_{13}\\
    C_{21}&C_{22}&C_{23}\\
    C_{31}&C_{32}&C_{33}\\
\end{matrix}\right]
\end{align}

Similiar to 2D model, we can using the Taylor expansion, $\mathbf{u}(\mathbf{x},t)=(u(\mathbf{x},t) , v(\mathbf{x},t), w(\mathbf{x},t))^T$ to obtain the tensor, where
 \begin{equation}
   \bm{\eta}=\mathbf{u}(\mathbf{x}^{\prime},t)-\mathbf{u}(\mathbf{x},t)=\sum_{n=1}^\infty\frac1{n!}\Bigg(\xi_1\frac\partial{\partial x_1}+\xi_2\frac\partial{\partial x_2}+\xi_3\frac\partial{\partial x_3}\Bigg)^n\mathbf{u}(\mathbf{x},t)
\end{equation}

Using (55) , we can obtain
\begin{equation}
P(\mathbf{x}, t)=\sum_{n=1}^\infty\frac1{n!}P_n(\mathbf{x}, t)\end{equation}
where
\begin{align}
P_n(\mathbf{x}, t)& =\frac{1}{\pi \delta^4}\ds \int_{\mathbf{\mathcal{B}_{\delta}(\mathbf{x})}}\dfrac{\mathbf{C}(\boldsymbol{\xi}\bullet\nabla)^n\mathbf{\bm{u}}}{\|\mathbf{\bm{\xi}}\|^{3}}dV_{\mathbf{\mathbf{x}^{\prime}}} \\
&=\frac{1}{\pi \delta^4}\ds \int_{\mathbf{\mathcal{B}_{\delta}(\mathbf{x})}}\dfrac{(\boldsymbol{\xi}\bullet\nabla)^n\begin{pmatrix}C_{11}u+
    C_{12}v+C_{13}w\\ C_{21}u+
C_{22}v+C_{23}w\\
C_{31}u+
C_{32}v+C_{33}w\end{pmatrix}}{\|\mathbf{\bm{\xi}}\|^{3}}dV_{\mathbf{\mathbf{x}^{\prime}}}
\end{align}
  then we obtain 
 {\smaller\begin{align}
 \lim_{\delta \to 0} P(\mathbf{x}, t)&=\lim_{\delta \to 0}\frac{1}{2\pi \delta^4}\ds \int_{\mathbf{\mathcal{B}_{\delta}(\mathbf{x})}}\dfrac{\mathbf{C}(\boldsymbol{\xi}\bullet\nabla)^{2}\mathbf{\bm{u}}}{\|\mathbf{\bm{\xi}}\|^{3}}dV_{\mathbf{\mathbf{x}^{\prime}}}=[P_x,P_y,P_z]^T,\end{align}}
 where
\begin{align}
P_x=&\left(\dfrac{D^{1\hspace{0.12em}1}_{.1.1}}{10}+\dfrac{D^{1\hspace{0.12em}2}_{.1.2}}{30}+\dfrac{D^{1\hspace{0.12em}3}_{.1.3}}{30}\right) u_{x x}+ \left(  \dfrac{D^{1\hspace{0.12em}1}_{.1.1}}{30}+\dfrac{D^{1\hspace{0.12em}2}_{.1.2}}{10}+\dfrac{D^{1\hspace{0.12em}3}_{.1.3}}{30}\right)  u_{x y}+\left( \dfrac{D^{1\hspace{0.12em}1}_{.1.1}}{30}+\dfrac{D^{1\hspace{0.12em}2}_{.1.2}}{30}+\dfrac{D^{1\hspace{0.12em}3}_{.1.3}}{10}\right) u_{z z}\notag\\
+&\left(\dfrac{D^{1\hspace{0.12em}1}_{.2.1}}{10}+\dfrac{D^{1\hspace{0.12em}2}_{.2.2}}{30}+\dfrac{D^{1\hspace{0.12em}3}_{.2.3}}{30}\right)v_{x x}+\left(\dfrac{D^{1\hspace{0.12em}1}_{.2.1}}{30}+\dfrac{D^{1\hspace{0.12em}2}_{.2.2}}{10}+\dfrac{D^{1\hspace{0.12em}3}_{.2.3}}{30}\right) v_{y y}+\left( \dfrac{D^{1\hspace{0.12em}1}_{.2.1}}{30}+\dfrac{D^{1\hspace{0.12em}2}_{.2.2}}{30}+\dfrac{D^{1\hspace{0.12em}3}_{.2.3}}{10}\right) v_{z z}\notag\\
+&\left(\dfrac{D^{1\hspace{0.12em}1}_{.3.1}}{10}+\dfrac{D^{1\hspace{0.12em}2}_{.3.2}}{30}+\dfrac{D^{1\hspace{0.12em}3}_{.3.3}}{30}\right)w_{x x}+\left( \dfrac{D^{1\hspace{0.12em}1}_{.3.1}}{30}+\dfrac{D^{1\hspace{0.12em}2}_{.3.2}}{10}+\dfrac{D^{1\hspace{0.12em}3}_{.3.3}}{30}\right)w_{y y}+\left( \dfrac{D^{1\hspace{0.12em}1}_{.3.1}}{30}+\dfrac{D^{1\hspace{0.12em}2}_{.3.2}}{30}+\dfrac{D^{1\hspace{0.12em}3}_{.3.3}}{10}\right) w_{z z} \notag\\
& +\dfrac{2D^{1\hspace{0.12em}2}_{.1.1}}{15}u_{x y}+\dfrac{2D^{1\hspace{0.12em}3}_{.1.1}}{15}u_{x z}+\dfrac{2D^{1\hspace{0.12em}3}_{.1.2}}{15} u_{y z}+\dfrac{2D^{1\hspace{0.12em}2}_{.2.1}}{15}v_{x y}+\dfrac{2D^{1\hspace{0.12em}3}_{.2.1}}{15}v_{x z}+\dfrac{2D^{1\hspace{0.12em}3}_{.2.2}}{15}v_{y z} \notag\\
& +\dfrac{2D^{1\hspace{0.12em}3}_{.2.1}}{15}w_{x y}+\dfrac{2D^{1\hspace{0.12em}3}_{.3.1}}{15}w_{x z}\dfrac{2D^{1\hspace{0.12em}3}_{.3.2}}{15}w_{y z},\notag\\
P_y=
& \left(\dfrac{D^{1\hspace{0.12em}1}_{.2.1}}{10}+\dfrac{D^{1\hspace{0.12em}2}_{.2.2}}{30}+\dfrac{D^{1\hspace{0.12em}3}_{.2.3}}{30}\right) u_{x x}+\left(\dfrac{D^{1\hspace{0.12em}1}_{.2.1}}{30}+\dfrac{D^{1\hspace{0.12em}2}_{.2.2}}{10}+\dfrac{D^{1\hspace{0.12em}3}_{.2.3}}{30}\right) u_{y y}+\left( \dfrac{D^{1\hspace{0.12em}1}_{.2.1}}{30}+\dfrac{D^{1\hspace{0.12em}2}_{.2.2}}{30}+\dfrac{D^{1\hspace{0.12em}3}_{.2.3}}{10}\right) u_{z z}\notag\\
+&\left(  \dfrac{D^{1\hspace{0.12em}1}_{.1.1}}{30}+\dfrac{D^{1\hspace{0.12em}2}_{.1.2}}{10}+\dfrac{D^{1\hspace{0.12em}3}_{.1.3}}{30}\right) v_{x x}+\left(\dfrac{D^{2\hspace{0.12em}1}_{.2.1}}{30}+\dfrac{D^{2\hspace{0.12em}2}_{.2.2}}{10}+\dfrac{D^{2\hspace{0.12em}3}_{.2.3}}{30}\right)v_{y y}+\left(\dfrac{D^{2\hspace{0.12em}1}_{.2.1}}{30}+\dfrac{D^{2\hspace{0.12em}2}_{.2.2}}{30}+\dfrac{D^{2\hspace{0.12em}3}_{.2.3}}{10}\right) v_{zz}\notag\\
+&\left( \dfrac{D^{2\hspace{0.12em}1}_{.3.1}}{10}+\dfrac{D^{2\hspace{0.12em}2}_{.3.2}}{30}+\dfrac{D^{2\hspace{0.12em}3}_{.3.3}}{30}\right)w_{x x}+\left( \dfrac{D^{2\hspace{0.12em}1}_{.3.1}}{30}+\dfrac{D^{2\hspace{0.12em}2}_{.3.2}}{10}+\dfrac{D^{2\hspace{0.12em}3}_{.3.3}}{30}\right)w_{y y}+\left( \dfrac{D^{2\hspace{0.12em}1}_{.3.1}}{30}+\dfrac{D^{2\hspace{0.12em}2}_{.3.2}}{30}+\dfrac{D^{2\hspace{0.12em}3}_{.3.3}}{10}\right) w_{z z}\notag\\
& +\dfrac{2D^{1\hspace{0.12em}2}_{.2.1}}{15} u_{x y}+\dfrac{2D^{1\hspace{0.12em}3}_{.2.1}}{15} u_{x z}+\dfrac{2D^{1\hspace{0.12em}3}_{.2.2}}{15} u_{y z}+\dfrac{2D^{2\hspace{0.12em}2}_{.2.1}}{15}v_{x y}+\dfrac{2D^{2\hspace{0.12em}3}_{.2.1}}{15} v_{x z}+\dfrac{2D^{2\hspace{0.12em}3}_{.2.2}}{15} v_{y z} \notag\\
& +\dfrac{2D^{1\hspace{0.12em}3}_{.2.2}}{15} w_{x y}\dfrac{2D^{1\hspace{0.12em}3}_{.3.2}}{15} w_{x z}+\dfrac{2D^{2\hspace{0.12em}3}_{.3.2}}{15} w_{y z},\notag\\
P_z=
&\left(\dfrac{D^{1\hspace{0.12em}1}_{.3.1}}{10}+\dfrac{D^{1\hspace{0.12em}2}_{.3.2}}{30}+\dfrac{D^{1\hspace{0.12em}3}_{.3.3}}{30}\right) u_{x x}+\left( \dfrac{D^{1\hspace{0.12em}1}_{.3.1}}{30}+\dfrac{D^{1\hspace{0.12em}2}_{.3.2}}{10}+\dfrac{D^{1\hspace{0.12em}3}_{.3.3}}{30}\right) u_{y y}+\left( \dfrac{D^{1\hspace{0.12em}1}_{.3.1}}{30}+\dfrac{D^{1\hspace{0.12em}2}_{.3.2}}{30}+\dfrac{D^{1\hspace{0.12em}3}_{.3.3}}{10}\right)u_{z z}\notag\\
+&\left( \dfrac{D^{2\hspace{0.12em}1}_{.3.1}}{10}+\dfrac{D^{2\hspace{0.12em}2}_{.3.2}}{30}+\dfrac{D^{2\hspace{0.12em}3}_{.3.3}}{30}\right)v_{x x}+\left( \dfrac{D^{2\hspace{0.12em}1}_{.3.1}}{30}+\dfrac{D^{2\hspace{0.12em}2}_{.3.2}}{10}+\dfrac{D^{2\hspace{0.12em}3}_{.3.3}}{30}\right)v_{y y}+\left( \dfrac{D^{2\hspace{0.12em}1}_{.3.1}}{30}+\dfrac{D^{2\hspace{0.12em}2}_{.3.2}}{30}+\dfrac{D^{2\hspace{0.12em}3}_{.3.3}}{10}\right)v_{z z}\notag\\
&+\left( \dfrac{D^{1\hspace{0.12em}1}_{.1.1}}{30}+\dfrac{D^{1\hspace{0.12em}2}_{.1.2}}{30}+\dfrac{D^{1\hspace{0.12em}3}_{.1.3}}{10}\right)w_{x x}+\left(\dfrac{D^{2\hspace{0.12em}1}_{.2.1}}{30}+\dfrac{D^{2\hspace{0.12em}2}_{.2.2}}{30}+\dfrac{D^{2\hspace{0.12em}3}_{.2.3}}{10}\right)w_{y y}+\left( \dfrac{D^{3\hspace{0.12em}1}_{.3.1}}{30}+\dfrac{D^{3\hspace{0.12em}2}_{.3.2}}{30}+\dfrac{D^{3\hspace{0.12em}3}_{.3.3}}{10}\right)w_{z z} \notag\\
& +\dfrac{2D^{1\hspace{0.12em}3}_{.2.1}}{15} u_{x y}+\dfrac{2D^{1\hspace{0.12em}3}_{.3.1}}{15} u_{x z}\dfrac{2D^{1\hspace{0.12em}3}_{.3.2}}{15} u_{y z}+\dfrac{2D^{1\hspace{0.12em}3}_{.2.2}}{15} v_{x y}\dfrac{2D^{1\hspace{0.12em}3}_{.3.2}}{15}v_{x z}+\dfrac{2D^{2\hspace{0.12em}3}_{.3.2}}{15}v_{y z}\notag \\
& +2\left( \dfrac{D^{1\hspace{0.12em}1}_{.2.1}}{30}+\dfrac{D^{1\hspace{0.12em}2}_{.2.2}}{30}+\dfrac{D^{1\hspace{0.12em}3}_{.2.3}}{10}\right)w_{x y}+2\left( \dfrac{D^{1\hspace{0.12em}1}_{.3.1}}{30}+\dfrac{D^{1\hspace{0.12em}2}_{.3.2}}{30}+\dfrac{D^{1\hspace{0.12em}3}_{.3.3}}{10}\right)w_{x z}+2\left( \dfrac{D^{2\hspace{0.12em}1}_{.3.1}}{30}+\dfrac{D^{2\hspace{0.12em}2}_{.3.2}}{30}+\dfrac{D^{2\hspace{0.12em}3}_{.3.3}}{10}\right)w_{y z}\notag.
\end{align}

Then we can compute the expression by three component, which is shown in Appendix A. then we can obtain the expression of tensor in 3D model:

\begin{equation}
  \begin{aligned}
     &\quad \resizebox{1\textwidth}{!}{%
     \(
    \mathcal{D}=\left[
     \begin{matrix}
     3(4Q_{41} - Q_{44} - Q_{55})&15Q_{14}&15Q_{55}&15Q_{14}&3(4Q_{44} - Q_{11} - Q_{55})&15Q_{45}&15Q_{55}&15Q_{45}&3(4Q_{55} - Q_{11} - Q_{44})  \\
     3(4Q_{14} - Q_{24} - Q_{56})&\frac{15}{2}(Q_{12} + Q_{44}) & \frac{15}{2}(Q_{16} + Q_{45}) & \frac{15}{2}(Q_{12} + Q_{44}) &3(4Q_{24} - Q_{14} - Q_{56}) &\frac{15}{2}(Q_{46} + Q_{45})&\frac{15}{2}(Q_{16} + Q_{45}) & \frac{15}{2}(Q_{46} + Q_{45})&3(4Q_{56} - Q_{24} - Q_{44})\\
     3(4Q_{15} - Q_{46} - Q_{55}) &\frac{15}{2}(Q_{16} + Q_{45}) &\frac{15}{2}(Q_{13} + Q_{55}) &\frac{15}{2}(Q_{16} + Q_{45}) &3(Q_{46} - Q_{15} - Q_{35}) &\frac{15}{2}(Q_{34} + Q_{56})& \frac{15}{2}(Q_{13} + Q_{55}) & \frac{15}{2}(Q_{34} + Q_{56})& 3(4Q_{35} - Q_{15} - Q_{46})\\
    3(4Q_{14} - Q_{24} - Q_{56})&\frac{15}{2}(Q_{12} + Q_{44}) & \frac{15}{2}(Q_{16} + Q_{45}) & \frac{15}{2}(Q_{12} + Q_{44}) &3(4Q_{24} - Q_{14} - Q_{56}) &\frac{15}{2}(Q_{46} + Q_{45})&\frac{15}{2}(Q_{16} + Q_{45}) & \frac{15}{2}(Q_{46} + Q_{45})&3(4Q_{56} - Q_{24} - Q_{44})\\
    3(4Q_{44} - Q_{22} - Q_{66}) & 15Q_{24} &15Q_{46} &15Q_{24} &3(4Q_{12} - Q_{44} - Q_{66}) & 15Q_{26}& 15Q_{46} & 15Q_{26}& 3(4Q_{66} - Q_{22} - Q_{44})\\
    3(4Q_{45} - Q_{26} - Q_{36}) &\frac{15}{2}(Q_{46} + Q_{45}) & \frac{15}{2}(Q_{34} + Q_{56}) &\frac{15}{2}(Q_{46} + Q_{45}) &3(4Q_{26} - Q_{45} - Q_{36}) &\frac{15}{2}(Q_{23} + Q_{66})&\frac{15}{2}(Q_{34} + Q_{56}) & \frac{15}{2}(Q_{23} + Q_{66}) &3(4Q_{36} - Q_{45} - Q_{26})\\
       3(4Q_{15} - Q_{46} - Q_{55}) &\frac{15}{2}(Q_{16} + Q_{45}) &\frac{15}{2}(Q_{13} + Q_{55}) &\frac{15}{2}(Q_{16} + Q_{45}) &3(Q_{46} - Q_{15} - Q_{35}) &\frac{15}{2}(Q_{34} + Q_{56})& \frac{15}{2}(Q_{13} + Q_{55}) & \frac{15}{2}(Q_{34} + Q_{56})& 3(4Q_{35} - Q_{15} - Q_{46})\\
           3(4Q_{45} - Q_{26} - Q_{36}) &\frac{15}{2}(Q_{46} + Q_{45}) & \frac{15}{2}(Q_{34} + Q_{56}) &\frac{15}{2}(Q_{46} + Q_{45}) &3(4Q_{26} - Q_{45} - Q_{36}) &\frac{15}{2}(Q_{23} + Q_{66})&\frac{15}{2}(Q_{34} + Q_{56}) & \frac{15}{2}(Q_{23} + Q_{66}) &3(4Q_{36} - Q_{45} - Q_{26})\\
       3(4Q_{55} - Q_{66} - Q_{33}) & 15Q_{56} & 15Q_{35}&  15Q_{56} &  3(4Q_{16} - Q_{55} - Q_{33}) &15Q_{36} & 15Q_{35} &15Q_{36}&3(4Q_{35} - Q_{55} - Q_{66})
    \end{matrix}
     \right]
    \)
    }
  \end{aligned}
 \end{equation}
 \begin{remark}
     Similar to the two-dimensional model, if the tensors in the three-dimensional model satisfy three symmetries, the elements in the elasticity matrix must satisfy relationship $Q_{11}=Q_{22}=Q_{33}$, $Q_{14}+Q_{16}+Q_{35}=0$, $Q_{25}+Q_{26}+Q_{36}=0$, $Q_{45}+Q_{46}+Q_{56}=0$. This results in the model constructed with this tensor being unsuitable for general anisotropic materials.
 \end{remark}

\subsection{The critical failure criterion of Ti-PD for 3D isotropic materials}
In this section, we examine the concept of failure in the 3D Ti-PD framework for isotropic materials.   $G_0$ and  $s_0$ for the 3D case of isotropic materials are defined as follows \cite{SA2005}:
 	\begin{equation}\label{3DG0}
 		\begin{aligned}
	 G_0&= \ds \int_{0}^{\delta}\int_{0}^{2\pi}\int_{z}^{\delta}\int_{0}^{arccos z/\xi} \frac{1}{2}\mathbf{f}(\bm{\eta},\bm{\xi})\cdot\bm{\eta}\xi^2 sin\phi d\phi d\xi d\theta dz=
     \frac{3Es_0^2\delta(3-2\nu)}{10(1-2\nu)(1+\nu)}
	\end{aligned}
\end{equation}
An explanation of this computation are shown in Fig. \ref{Fig:s0}. 
 	\begin{figure}[ht]
		\centering            
		\includegraphics[scale=0.5]{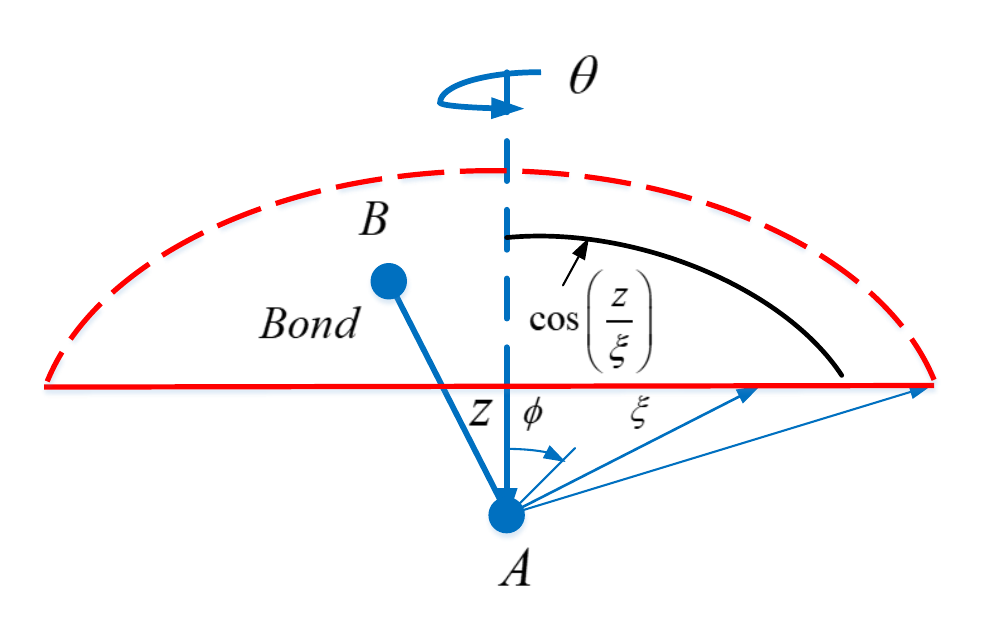}  
		\caption{Evaluation of fracture energy $G_0$  in 3D model for each material point $\mathbf{A}$. Here red real line is the fracture surface.  $\mathbf{A}$ and  $\mathbf{B}$ are material points and  $0\leq z \leq \delta$. The energy required to break the bond between material point $\mathbf{A}$ and material point $\mathbf{B}$ in a spherical coordinate system can be derived through equation \eqref{3DG0}.} 
		\label{Fig:s03d}
	\end{figure}
Thus
\begin{equation}\label{3DOSB-PDs0}
	s_0=\sqrt{\frac{10G_0(1-2v)(1+v)}{3E\delta(3-2v)}}
\end{equation}
More generally, when Poisson's ratio is $1/4$, the expression of $s_0$ can be obtained by (\refeq{3DOSB-PDs0}) as:
\begin{equation}\label{3DBB-PDs0}
	s_0=\sqrt{\frac{5 G_0}{6E\delta}}
\end{equation}
which means that the new $s_0$ is equal to $s_0$ in the three-dimensional BB-PD model when simulate isotropic materials.
\begin{remark}
For 3D OSB-PD, the $s_0$ can be expressed as:
\begin{equation}
     s_0=
 \sqrt{\frac{G_0}{\left(3 \mu+\left(\frac{3}{4}\right)^4\left(\kappa-\frac{5 \mu}{3}\right)\right) \delta}} =\sqrt{\frac{1536(1+\nu)(1-2\nu)G_0}{(2061-3636\nu)E\delta}}
\end{equation}
Similar to the 2D models, the fracture criteria for OSB-PD and Ti-PD are consistent only when the Poisson's ratio is equal to 1/4. They both turn to (\refeq{3DBB-PDs0}). A comparison will be provided in Section 5.3.
\end{remark}
\section{Numerical examples}
In this section, we provide a series of numerical examples to illustrate the accuracy of the Ti-PD model across various scenarios. All experiments were carried out using MATLAB $^\circledR$ on a workstation featuring an Intel Xeon Gold 6240 CPU operating at 2.6 GHz, coupled with 2048 GB of installed memory.
\subsection{Validation}
In this subsection, we construct a continuous displacement function, which is substituted into the steady-state CCM equation to derive the corresponding load terms and boundary conditions, thereby formulating a steady-state elasticity problem. This problem is subsequently addressed using both the proposed and traditional models, with numerical solutions compared to the constructed displacement function. This comparison serves to evaluate the effectiveness of Ti-PD.

\subsubsection{Validation for 2D problems}
We begin by considering a 2D plate subject to Dirichlet boundary conditions. The region is defined as $\Omega=[-0.25,0.25]\times[-0.25,0.25]$,  
and the analytical displacement solution can be obtained by $\mathbf{u}(x,y)=(sin\pi(x+y),cos\pi(x+y))$. Here the voigt  matrix of elastic tensor $[\boldsymbol{\mathcal { C }}]$ is chosen as:
\begin{equation}
[\boldsymbol{\mathcal { C }}]
=
\left[
\begin{matrix}
    200&80&50\\
    80&150&40\\
    50&40&100\\
\end{matrix}
\right]
\end{equation}
Thus, the boundary conditions and right-hand-side loading terms are defined accordingly. The proposed tensor-involved peridynamic model (Ti-PD) and the traditional non-local bond-based peridynamic model (NOSB-PD) are employed to simulate this problem, with  $\delta =3\Delta x$ and a spatial discretization of $\Delta x=0.005$. 
 
\begin{figure}[htbp]
			\centering   
			\includegraphics[scale=0.38]{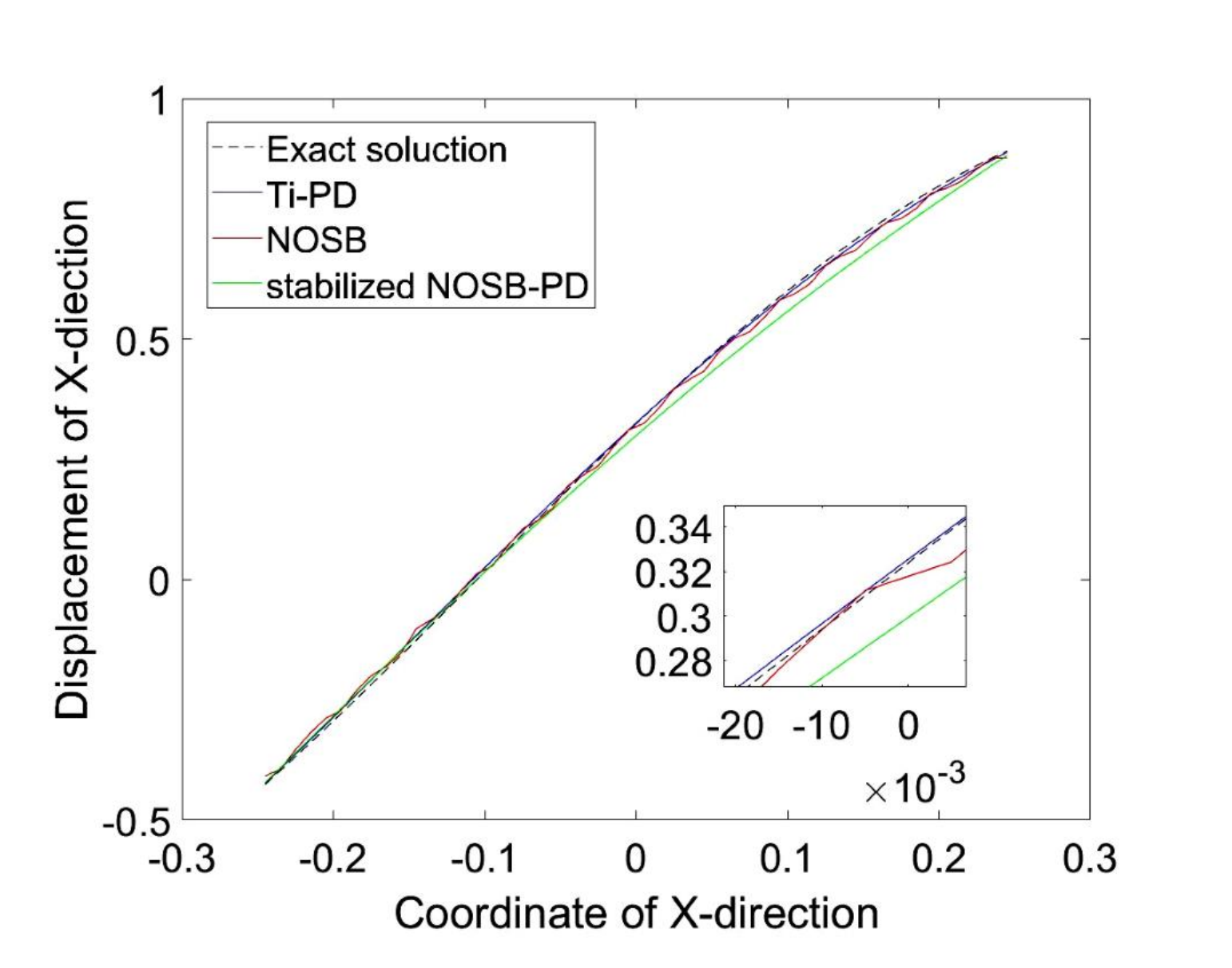} 
	\caption{Displacement variations in in x-direction at the line $(x,y=0)$ obtained by different models.}
	\label{va}
\end{figure}

The spatial discretization is implemented using the widely recognized meshfree method \cite{SA2005}. The detailed stiffness matrix is shown in Appendix A. Various models are utilized to assess accuracy in the displacement along the x-direction, including Ti-PD, bond-based peridynamic model (NOSB-PD), stabilized NOSB-PD as presented in \cite{Wa2019}, with the numerical results from Ti-PD illustrated in Fig.\ref{va}.

It can be observed that, compared to NOSB-PD, the Ti-PD model achieves stable solutions consistent with those obtained using correction methods, validating the accuracy of our proposed model for two-dimensional problems.

\subsubsection{Validation for 3D problems}
In this subsection, we consider a 3D block subject to Dirichlet boundary conditions. In this example, we choose the region as $\Omega=[-0.25,0.25]\times[-0.25,0.25]\times[-0.25,0.25]$ and set analytical solutions as $\mathbf{u}(x,y,z)=(sin\pi(x+y+z),sin\pi(x+y+z),cos\pi(x+y+z))$ .
The elastic matrix $[\boldsymbol{\mathcal { C }}]_0$  can be taken as:
\begin{equation}
[\boldsymbol{\mathcal { C }}]=\left(\begin{array}{cccccc}
230 &45&55 & 10 & 20 &15 \\
45&210&50&12&18& 10 \\
55&50&250&14&22 &16 \\
10 &12 &14 &90&25&18 \\
20 &18 &22&25&95& 20 \\
15& 10 &16&18& 20 &85
\end{array}\right)
\end{equation}
Thus, the boundary conditions and right-hand-side loading terms are defined accordingly.

\begin{figure}[htbp]
			\centering   
			\includegraphics[scale=0.22]{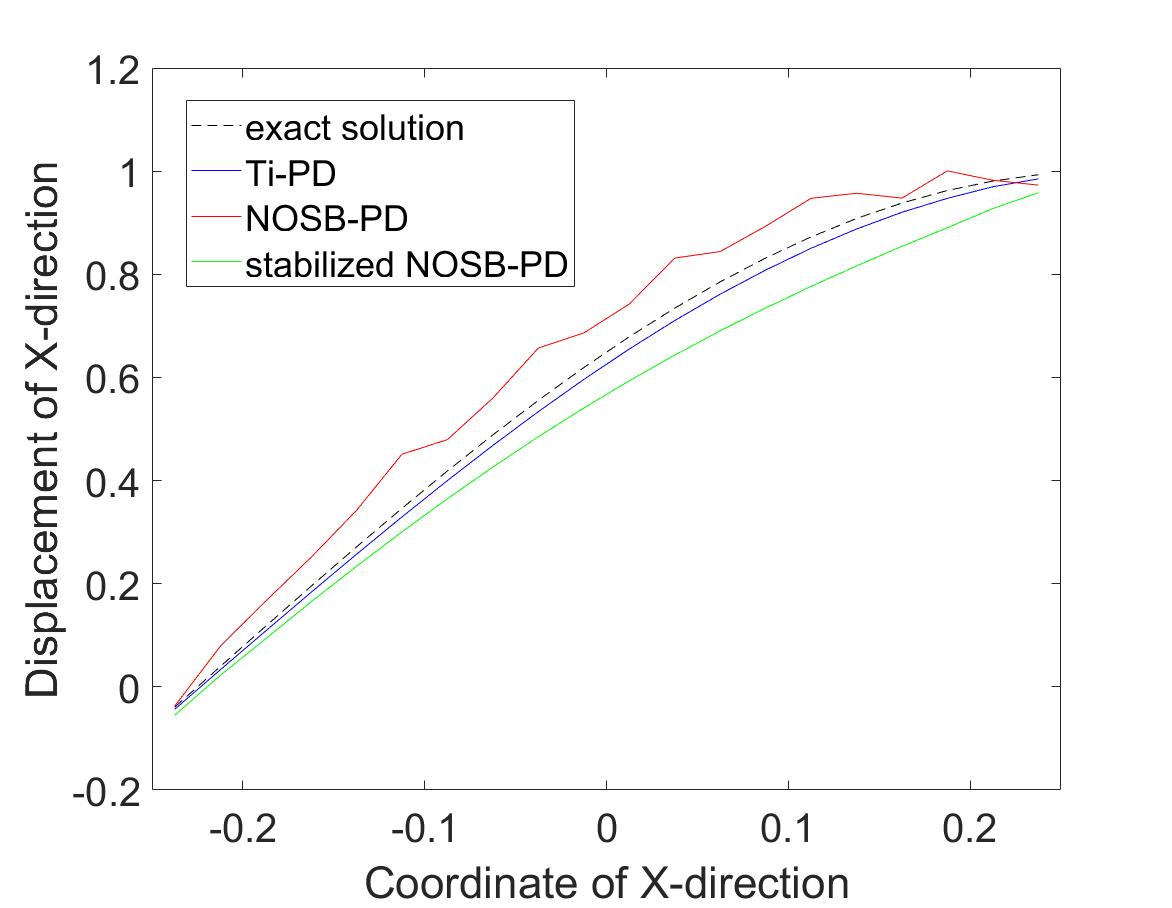} 
\caption{Displacement variations in x- direction at the line $(x,y=0,z=0)$ obtained by different models.}
	\label{va3D}
\end{figure}

The proposed  Ti-PD and traditional NOSB-PD is used to simulate this model, where   $\delta /\Delta x=3$ and the grid size $\Delta x=0.025$. Here the spatial discretization is chosen as meshfree method, The numerical result obtained by various models, including Ti-PD, NOSB, are shown in Fig.\ref{va3D}. 

In Fig. \ref{va3D}, several approaches were employed to obtain the displacement along  the x-direction. 
The first approach is the simulation results obtained by the Ti-PD, indicated by a solid yellow line; The second method is the NOSB-PD, depicted by a solid red line;The exact solution is depicted by a black dashed line.

As expected, the proposed Ti-PD model produced stable displacement solutions consistent with the actual displacement solutions. The direct application of NOSB-PD will lead to serious numerical oscillation.  This confirms the applicability of our proposed model for three-dimensional problems.

\subsection{Tensile simulation of plate with a circular hole}
\begin{figure}[ht]
	\centering            
	\includegraphics[scale=0.45]{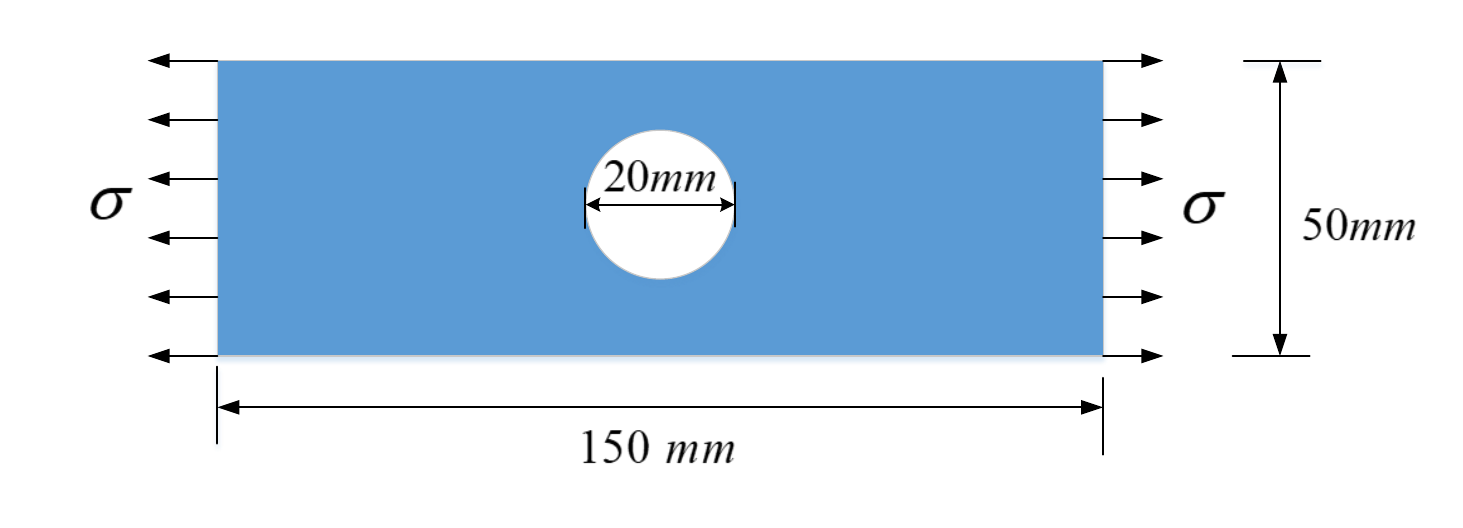}  
	\caption{Geometric dimensions of the plate with a circular hole. } 
	\label{yuankong}
\end{figure}

To illustrate the performance of the Ti-PD model, the displacement field of a plate with a circular hole is simulated. Both isotropic and anisotropic materials are considered in the subsequent subsections. As depicted in Fig.\ref{yuankong}, the plate measures 150 mm in length, 50 mm in width, and 1 mm in thickness, with a central hole diameter of 20 mm. The plate is subjected to displacement-controlled loading over 4000 time steps. The loading rate is set at  $1\times10^{-4}$mm/s, with a time step of 1.0 s. An adaptive dynamic relaxation method is employed to solve this quasi-static problem.

\subsubsection{Simulation for isotropic plates}
We conduct numerical experiments on isotropic plates, where the material properties are set at $E$=210 GPa and $\nu$=0.25. The analysis assumes plane stress conditions. The plate is uniformly discretized into 7,184 particles, featuring a node spacing of $\Delta x = 1 \, \text{mm}$ and $\delta = 3 \Delta x$.
%

\begin{figure}[htbp]
	\centering 	
			\centering   
			\includegraphics[scale=0.18]{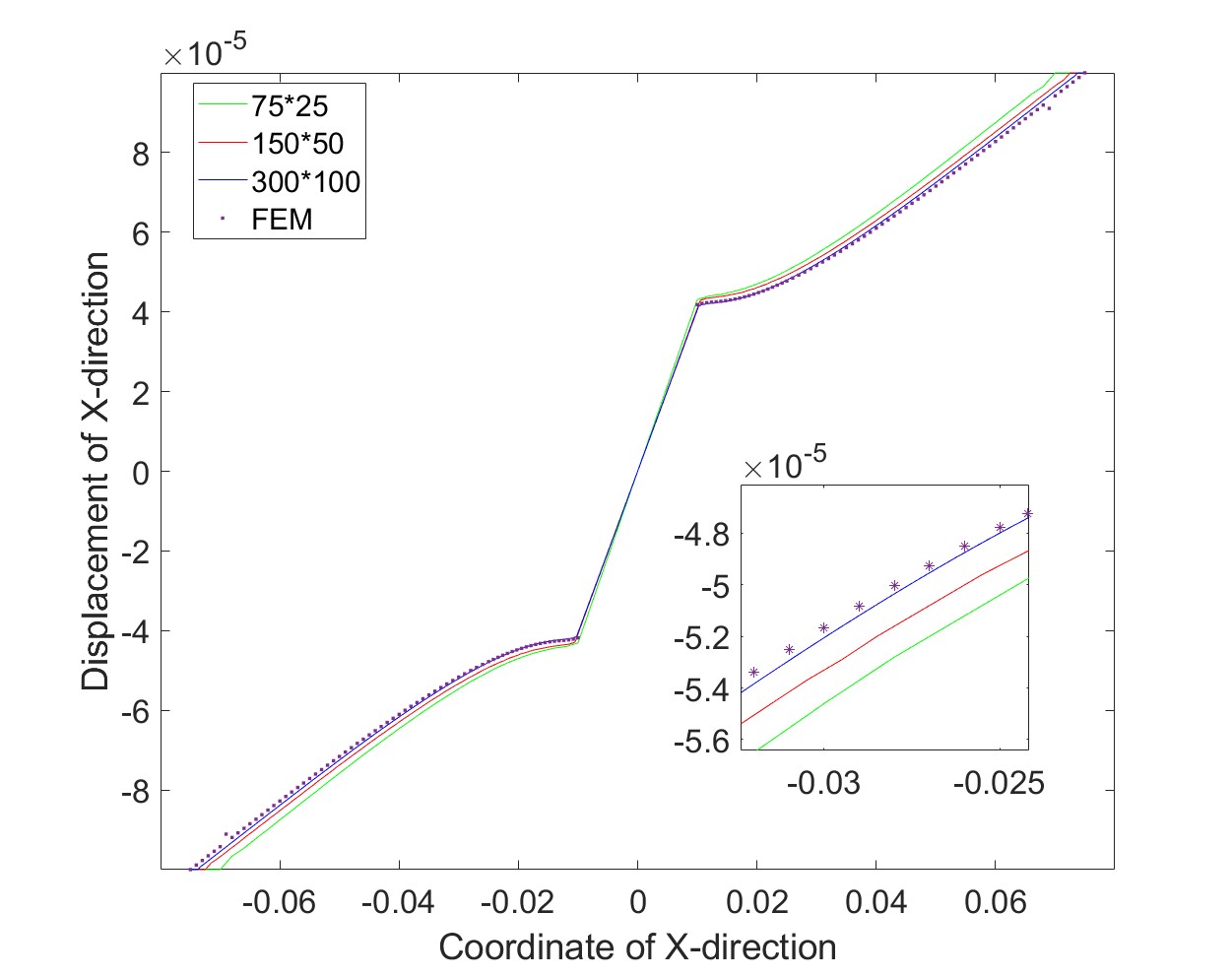} 
	\caption{Variation of horizontal displacement $u(x,y=0)$ for a plate with a circular hole obtained using the finite element method (FEM) and the Ti-PD model under different node numbers.}
	\label{yuankongtong}
\end{figure}

The meshfree method is employed for spatial discretization, with the number of material points selected as $75\times25$, $150\times50$, $300\times100$.   To validate the accuracy of the proposed model, we compared the Ti-PD solution with the finite element solution, and all results are illustrated in Fig.\ref{yuankongtong}.
As the number of nodes increases, the simulation outcomes increasingly converge with those obtained from the finite element method (FEM). For isotropic problems, the Ti-PD model consistently provides results that closely align with FEM findings. This strong correlation underscores the robustness of the Ti-PD model in accurately simulating material behavior under various loading conditions. Moreover, as the number of material points rises, the Ti-PD solution converges towards the FEM solution, demonstrating both the model's accuracy and its favorable convergence properties. This convergence further supports the assertion that the Ti-PD model effectively captures the underlying physical phenomena, positioning it as a reliable alternative to traditional methods.
\subsubsection{Simulation for anisotropic plates}
In this section, we investigate a plate with a circular hole composed of anisotropic material. The plate is a symmetric angle ply composite
 laminate of four graphite-epoxy laminae, with the following material properties are defined as follows: the elastic modulus in the fiber direction is  $E_{1} = 144.8 \, \text{GPa}$, the elastic modulus in the transverse direction is $E_{2} = 11.7 \, \text{GPa}$, the in-plane Poisson’s ratio $\nu_{12} = 0.21$, and the in-plane shear modulus $G_{12} = 9.66 \, \text{GPa}$. The unrotated material constitutive matrix $[\boldsymbol{\mathcal{C}}]$ in Voigt notation is calculated as:
\begin{equation}
[\boldsymbol{\mathcal { C }}]_0=\dfrac{1}{1-\nu_{12}\nu_{21}}\left[\begin{matrix}
        E_1&\nu_{21}E_2&0\\
        \nu_{21}E_2&E_2&0\\
        0&0&G_{12}(1-\nu_{12}\nu_{21})
    \end{matrix}\right]
\end{equation}
with $\nu_{21}E_2=\nu_{12}E_1$.   The stress at the bond is expressed in terms of a local coordinate system using the rotation matrix $R(\theta)$, which depends on the fibre orientation $\theta$ and is defined as
\begin{equation}
     R=\left[\begin{array}{ccc}
\cos ^2 \theta & \sin ^2 \theta & -2 \sin \theta \cos \theta \\
\sin ^2 \theta & \cos ^2 \theta & 2 \sin \theta \cos \theta \\
\sin \theta \cos \theta & -\sin \theta \cos \theta & \cos ^2 \theta-\sin ^2 \theta
\end{array}\right],
\end{equation}
where $\theta$ is the fiber angle,  with a selection range from $15^{\circ}$ to $90^{\circ}$. Then the rotated elastic matrix can be expressed as
\begin{equation}
    [\boldsymbol{\mathcal { C }}]=R[\boldsymbol{\mathcal { C }}]_0R^T.
\end{equation}

The analysis is performed under the assumption of plane stress conditions. The plate is uniformly discretized into 7,184 particles, featuring a grid spacing of $\Delta x = 1 \, \text{mm}$ and $\delta = 3 \Delta x$. 

\begin{figure}[htbp]
	\centering 	
    \subfigure[$\theta$=15$^{\circ}$]
    	{
		\begin{minipage}{7cm}
			\centering   
			\includegraphics[scale=0.15]{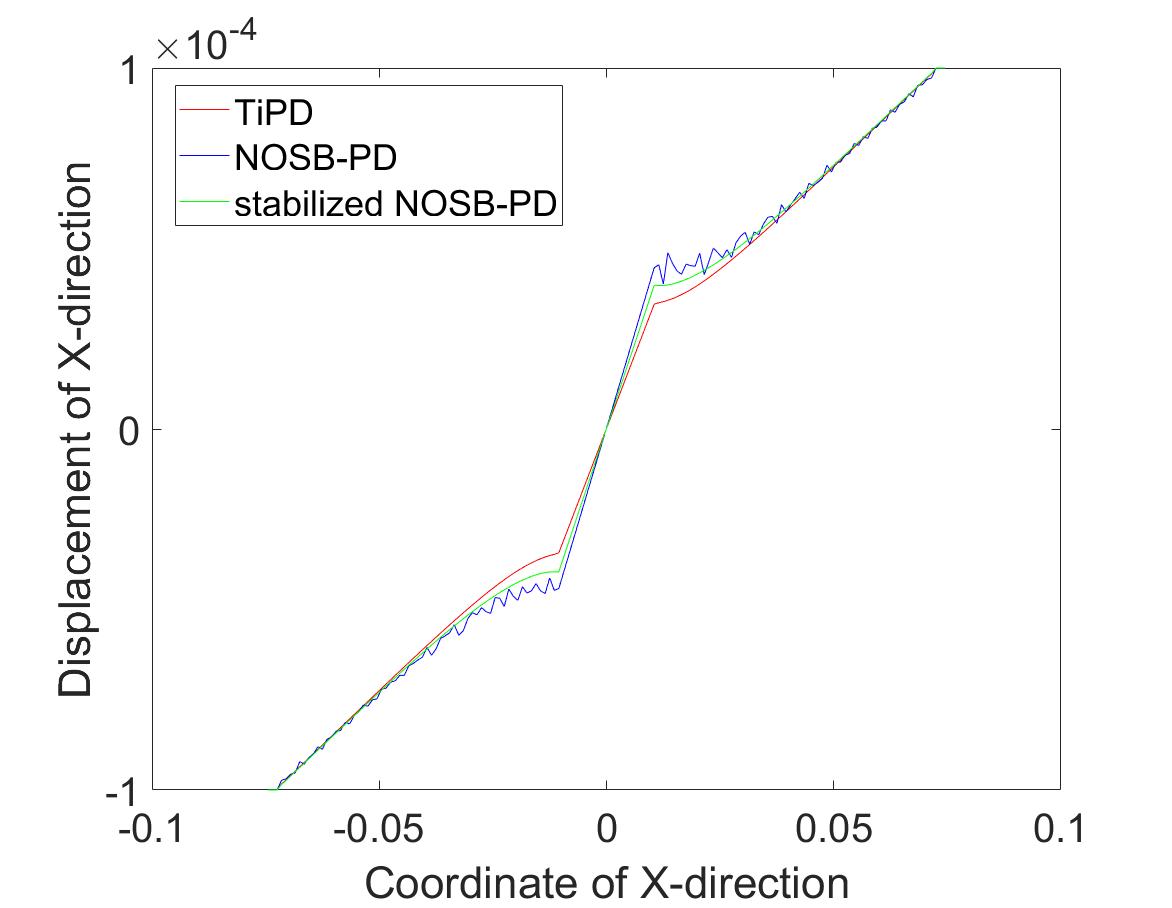} 
		\end{minipage}
	}
    \subfigure[$\theta$=30$^{\circ}$]
	{
		\begin{minipage}{7cm}
			\centering   
			\includegraphics[scale=0.15]{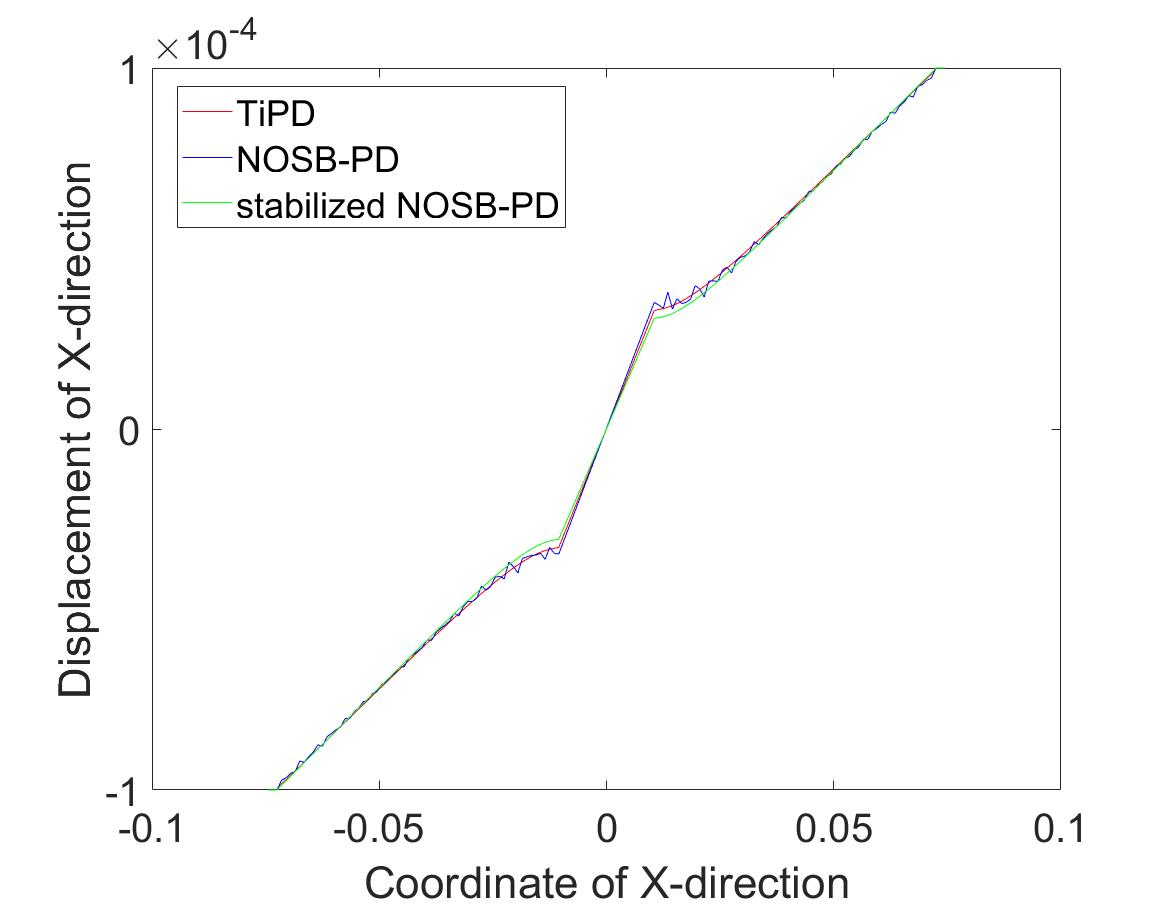} 
		\end{minipage}
	}	
    \subfigure[$\theta$=45$^{\circ}$]
       {
		\begin{minipage}{7cm}
			\centering   
			\includegraphics[scale=0.15]{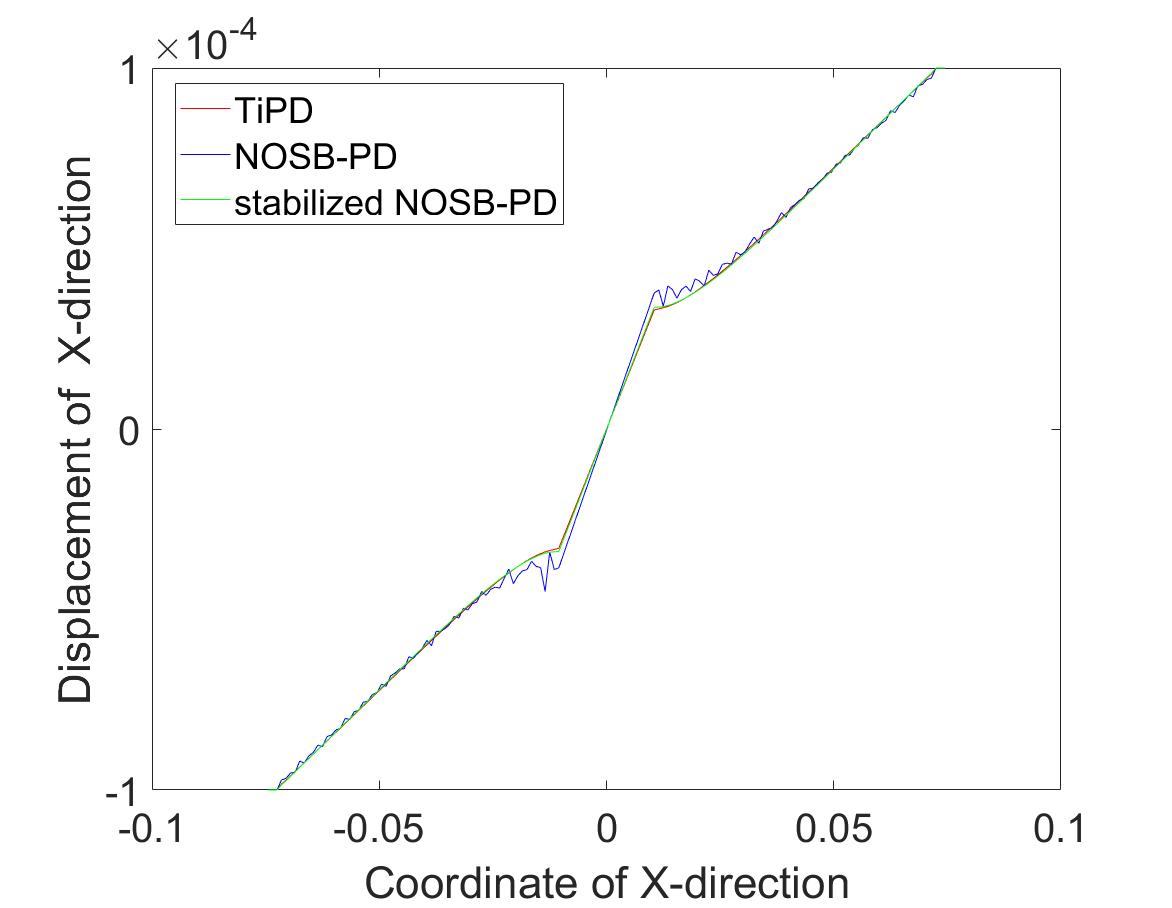} 
		\end{minipage}
	}	
    \subfigure[$\theta$=60$^{\circ}$]
        {
		\begin{minipage}{7cm}
			\centering   
			\includegraphics[scale=0.15]{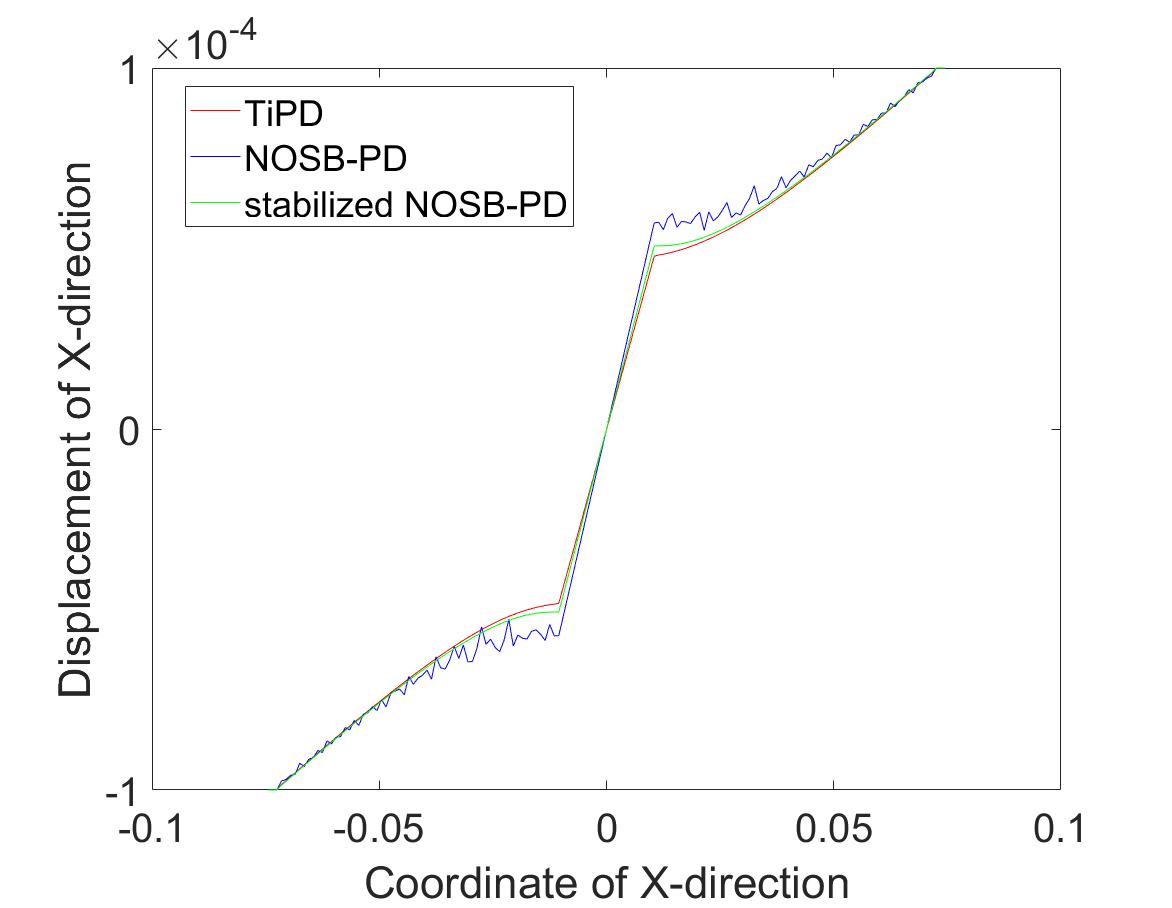} 
		\end{minipage}
	}
    \subfigure[$\theta$=75$^{\circ}$]
              {
		\begin{minipage}{7cm}
			\centering   
			\includegraphics[scale=0.15]{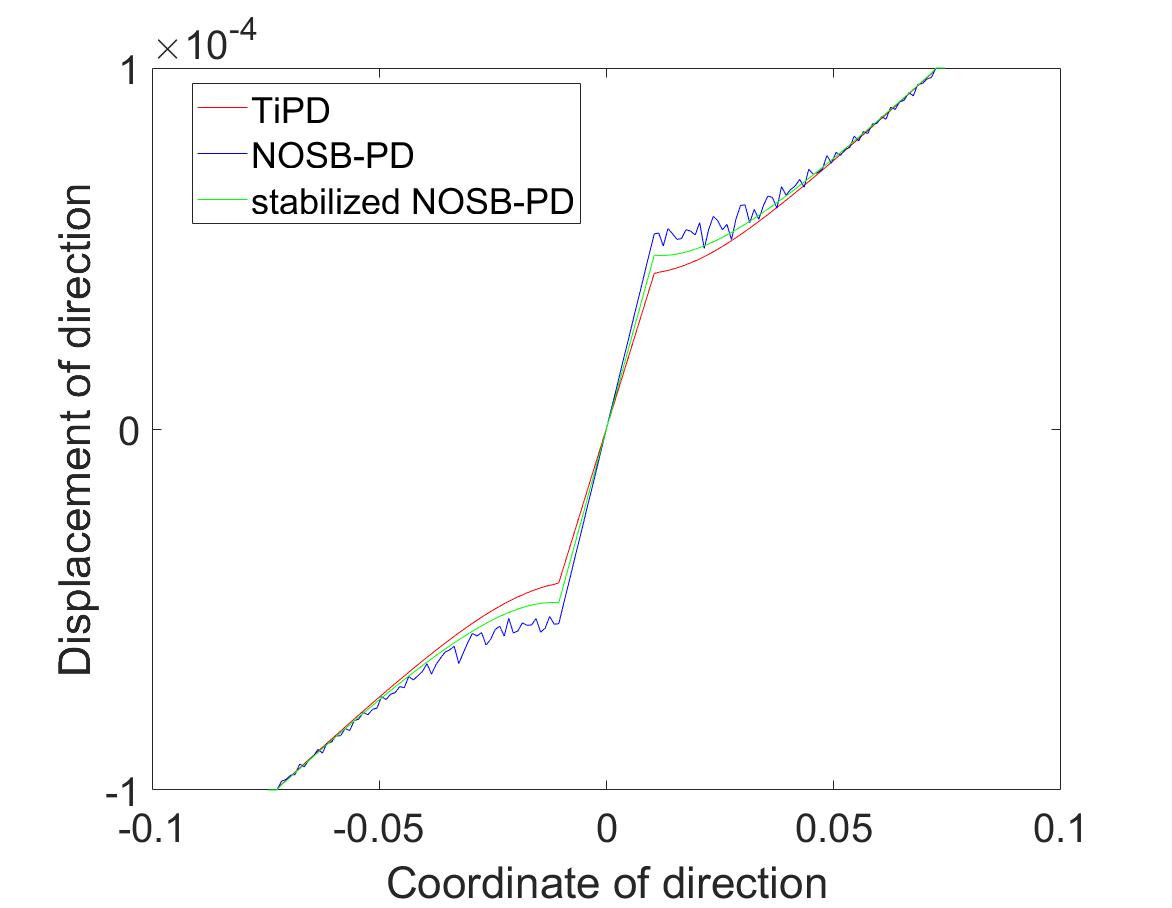} 
		\end{minipage}
	}	
    \subfigure[$\theta$=90$^{\circ}$]
        {
		\begin{minipage}{7cm}
			\centering   
			\includegraphics[scale=0.15]{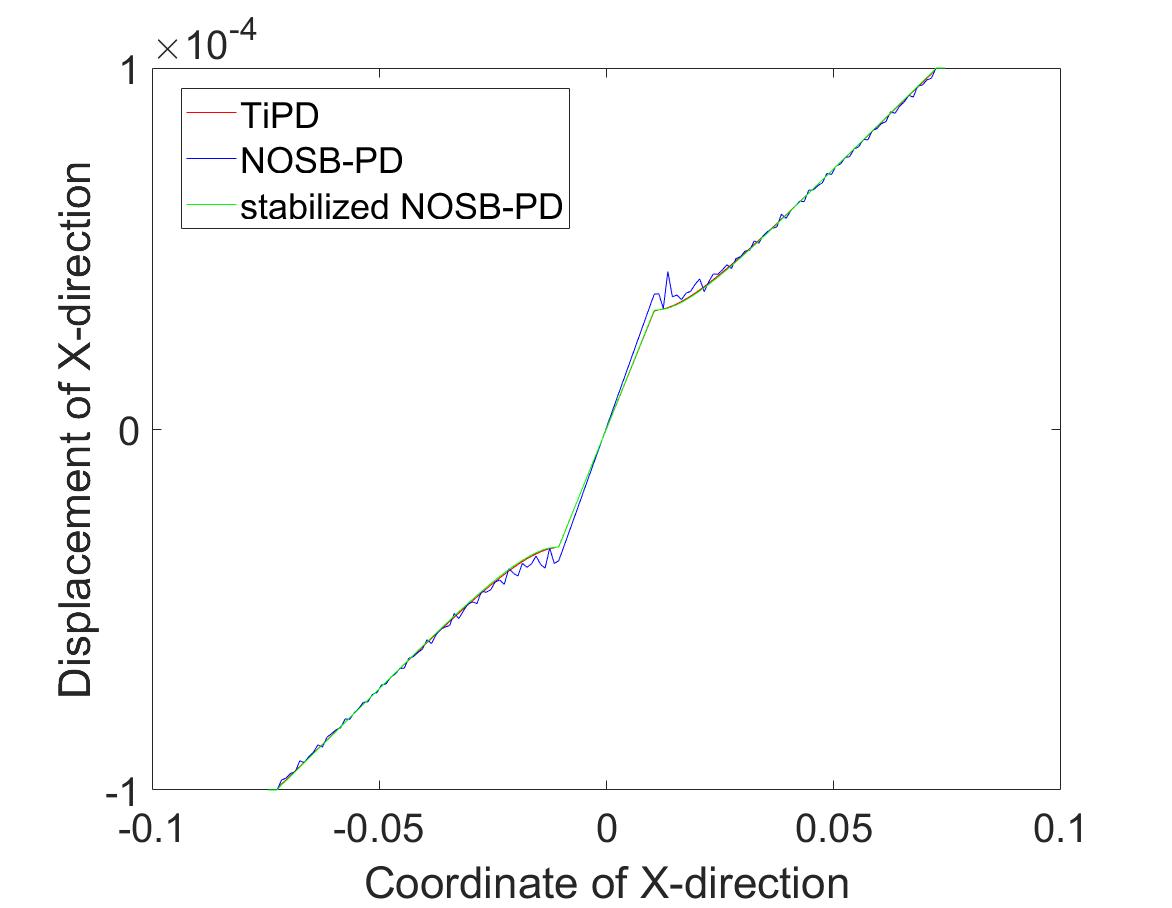} 
		\end{minipage}
	}
	\caption{Horizontal displacement $u(x,y=0)$ variation of plate with a circular hole in different fiber orientation angle.}
	\label{yuankongyi}
\end{figure}

To validate the accuracy of the models, several methods were utilized, with the displacement along the x-direction depicted in Fig. \ref{yuankongyi}. The frist model shows the simulation results obtained by the Ti-PD, indicated by a red dashed line. The second model is the NOSB-PD without any stabilization method, depicted by a solid blue line. The third model incorporates the NOSB model with the correction method proposed by \cite{Wa2019}, represented by a green line. 

Notably, the NOSB-PD model without correction exhibits significant numerical instabilities, particularly pronounced near the hole. This observation highlights the critical need for effective control of zero-energy modes to ensure accurate simulations. In contrast, the Ti-PD model demonstrates a remarkable ability to mitigate these instabilities, providing enhanced stability and accuracy in capturing the displacement field. This capability underscores the robustness of the Ti-PD model, especially in dynamic scenarios where precision is paramount. By effectively managing zero-energy modes, the Ti-PD model not only enhances computational reliability but also ensures that the predicted displacement behavior aligns consistently with the expected physical responses.

Furthermore, we calculated the CPU times for the various models when applied to anisotropic materials with different quantities of material points and chosen the material when when fiber orientation angle is 15 $^{\circ}$ as example, as summarized in Table 1. The results demonstrate that the Ti-PD model exhibits significantly greater computational efficiency compared to the other methods. This enhancement in efficiency can be attributed to the intrinsic structure of the bond-based framework utilized in Ti-PD, which is founded on a single integral formulation. In contrast, conventional state-based models necessitate a double integral approach, leading to increased computational overhead. Additionally, the stabilization formats in these conventional models often require the incorporation of correction terms, further exacerbating the computational burden. As a result, the Ti-PD model not only enhances accuracy but also significantly reduces the computational time required for large-scale problems, making it a highly efficient alternative in the analysis of crack propagation and material behavior.
\begin{table}
\begin{center}
    \renewcommand\arraystretch{1.3}
    \caption{CPU times computed by different models on plate with anisotropic material when fiber orientation angle is 30 $^{\circ}$.}
    \vspace{0.15in}
    \setlength{\tabcolsep}{3mm}{
        \begin{tabular}{l|l|ccc} 
            \hline
            \multicolumn{2}{c|}{Model} & NOSB-PD & Stabilized NOSB-PD & Ti-PD \\ 
            \hline
            \multirow{3}{*}{Nodes} & $300\times100$ &6m37s & 19m50s& 4m24s \\  
            \cline{2-5} 
            & $600\times200$ &40m40s&1h26m &20m49s\\    
            \cline{2-5} 
            & $1200\times400$ &3h20m& 6h19m & 1h25m\\ 
            \hline
        \end{tabular}
    }\label{tab:6}
\end{center}
\end{table}

\subsection{Tensile simulation of plate with a pre-existing crack}
\begin{figure}[ht]
 	\centering            
 	\includegraphics[scale=0.6]{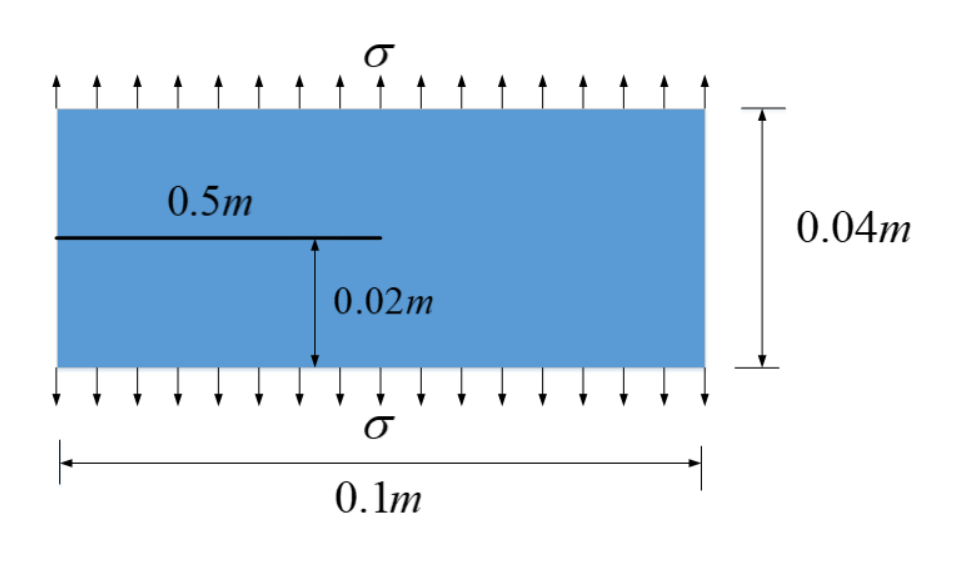}  
 	\caption{Geometry and boundary conditions of the plate with a pre-existing crack. } 
 	\label{boli}
 \end{figure}

To evaluate the performance of the Ti-PD model To assess the performance of the Ti-PD model in simulating crack propagation, we conduct numerical simulations of crack branching in a thin rectangular plate with pre-existing notches, as depicted in Fig. \ref{boli}. A uniform tensile stress of $\sigma=12Mpa$ is applied to the edges of the plate, which is constructed from isotropic soda-lime glass. The mechanical properties of the material are as follows: $\rho=2440kg/m^3$ (Density), $E=72Gpa$ (Young’s modulus), $\nu=0.22$ (Poisson ratio), $G_0=135J/m^2$ (Fracture energy).

 
For the simulations,we set  \( \delta = 3 \Delta x \), with \( \Delta x = 2.5 \times 10^{-3} \, \text{m} \). Time discretization is carried out using the velocity Verlet algorithm, with physical time \( T = 4.6 \times 10^{-6} \, \text{s} \) and \( \Delta t = 2.5 \times 10^{-9} \, \text{s} \).

 \begin{figure}[ht]
	\centering    	
	\subfigure[the experimental result]
	{
		\begin{minipage}{7cm}
			\centering          
			\includegraphics[scale=0.6]{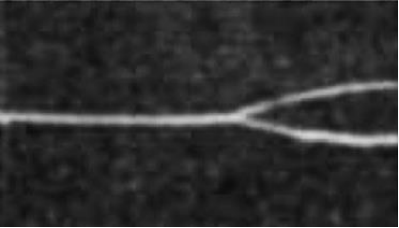}  
		\end{minipage}
	}	
	\subfigure[OSB-PD]
	{
		\begin{minipage}{7cm}
			\centering      
			\includegraphics[scale=0.6]{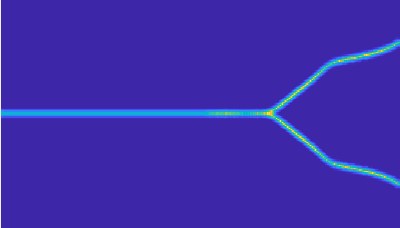}   
		\end{minipage}
	}	
 	\subfigure[Ti-PD with traditional $s_0$]
	{
		\begin{minipage}{7cm}
			\centering          
			\includegraphics[scale=0.6]{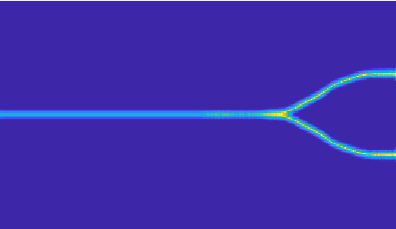}  
		\end{minipage}
	}	
	\subfigure[Ti-PD with novel $s_0$]
	{
		\begin{minipage}{7cm}
			\centering      
			\includegraphics[scale=0.6]{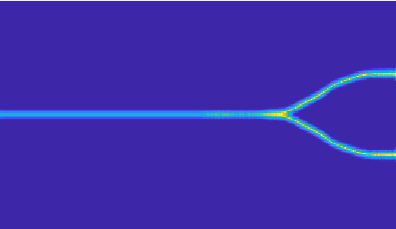}   
		\end{minipage}
	}	
	\caption{Crack propagation paths in the plate under the applied loading  $\sigma$=12 MPa obtained by various models, where the number of material points is chosen as $400\times160$: (a) the  experimental result in \cite{Lie2022}, (b) OSB-PD with traditional $s_0$, (c) Ti-PD with traditional $s_0$ (d) Ti-PD with novel $s_0$.
} 
	\label{12mpa}
\end{figure}

\begin{equation}
s_0==\sqrt{\frac{9\pi^2(1+\nu)(1-\nu)G}{(8(3\nu-1)+27\pi(1-\nu))E\delta}}
\end{equation}
Figs. \ref{12mpa} presents four images: a) results from physical experiments, b) simulation outcomes from the OSB-PD model utilizing the traditional critical stretch rate, c) simulation results from the Ti-PD model with the conventional $s_0$,and d) simulation results from the Ti-PD model employing a novel critical stretch rate.
  \begin{equation}
 	s_0=\sqrt{\frac{8\pi G_0(1-v^2)}{3E\delta(5+v)}}.
 \end{equation}
 Notably, the OSB-PD model predicts a larger crack angle, whereas the Ti-PD model yields results that align closely with the experimental results. This comparison highlights the enhanced accuracy of the Ti-PD model in effectively capturing crack propagation when contrasted with the OSB-PD model.

Another point to note is that we also provide simulation results for the Ti-PD model using the critical bond stretch value \(s_0\) derived from the traditional OSB-PD model. The results demonstrate that, regardless of whether \(s_0\) is derived from the OSB-PD or Ti-PD model, the Ti-PD model consistently produces more accurate crack angles. This suggests that the Ti-PD model can achieve improved precision in simulating displacements for discontinuous problems without requiring the introduction of new crack propagation criteria. Notably, this level of accuracy underscores the robustness of Ti-PD in capturing the behavior of crack paths across a variety of boundary and loading conditions, offering a substantial improvement over OSB-PD in dynamic fracture problems. By avoiding the complexities of additional criteria, Ti-PD maintains computational efficiency while enhancing precision in predicting material failure.

To validate the precision and robustness of the Ti-PD under higher loads, we kept the spatial grid size and time step constant while increasing the applied load to \( \sigma = 24 \) MPa. The physical time was extended to \( T = 3.2 \times 10^{-5} \) s, with 1500 time steps used to simulate the crack propagation. 

 \begin{figure}[ht]
	\centering    	
	\subfigure[the experimental result]
	{
		\begin{minipage}{7cm}
			\centering          
			\includegraphics[scale=0.6]{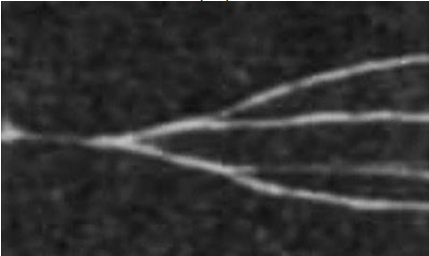}  
		\end{minipage}
	}	
	\subfigure[OSB-PD]
	{
		\begin{minipage}{7cm}
			\centering      
			\includegraphics[scale=0.6]{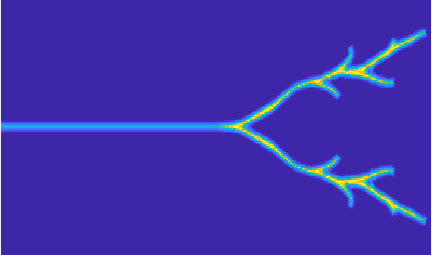}   
		\end{minipage}
	}	
 	\subfigure[Ti-PD with traditional $s_0$]
	{
		\begin{minipage}{7cm}
			\centering          
			\includegraphics[scale=0.34]{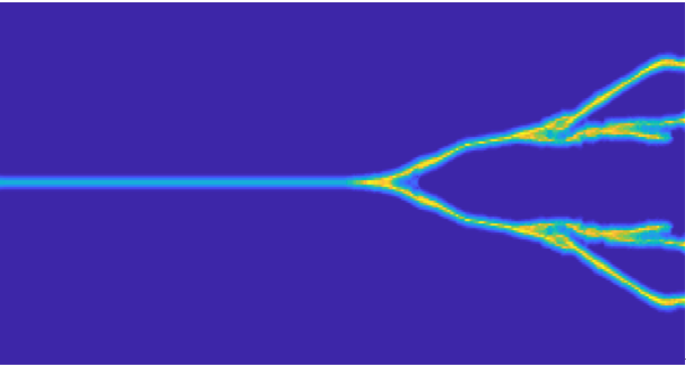}  
		\end{minipage}
	}	
	\subfigure[Ti-PD with novel $s_0$]
	{
		\begin{minipage}{7cm}
			\centering      
			\includegraphics[scale=0.6]{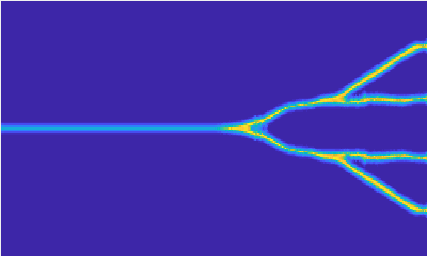}   
		\end{minipage}
	}	
	\caption{Crack propagation paths of the plate with the applied loading $\sigma$=12 MPa obtained by various models, where the number of material points is chosen as $400\times160$: (a) the experimental results in \cite{Lie2022}, (b) OSB-PD with traditional $s_0$, (c) Ti-PD with traditional $s_0$ (d) Ti-PD with novel $s_0$.
} 
	\label{24mpa}
\end{figure}

The results, presented in Fig. \ref{24mpa}, further highlight the limitations of the OSB-PD model. While OSB-PD exhibits fewer crack branches and generates erroneous secondary cracks, the proposed Ti-PD model produces crack branching patterns that closely align with experimental observations. This suggests that, despite retaining the simplicity of the bond-based peridynamic framework, Ti-PD is more adept at capturing the complex mechanisms involved in crack branching. The fidelity of Ti-PD to experimental data, especially in scenarios involving high tensile loads, reinforces its applicability in modeling anisotropic materials, which are often challenging for conventional bond-based approaches. This improvement opens up new avenues for simulating real-world fracture scenarios more accurately without needing more computationally demanding state-based models.

When the applied load is further increased, a noteworthy observation is that the traditional crack detection criteria begin to lose effectiveness. As shown in Figs. \ref{24mpa}(c) and (d), the Ti-PD model, when paired with the newly proposed crack detection criteria, produces results that align more closely with physical experiments, avoiding the formation of extraneous cracks. The old crack detection method, however, leads to the generation of additional cracks that are inconsistent with experimental observations. This suggests that the new criteria are more attuned to the actual fracture mechanics, especially under high-stress conditions. Despite this, it is important to emphasize that, regardless of which crack detection method is used, the Ti-PD model performs significantly better than the OSB-PD model. The absence of small, unphysical cracks in Ti-PD further solidifies its utility for accurately simulating crack growth in complex scenarios. This underscores the versatility of Ti-PD in handling anisotropic materials and large deformation problems, where traditional models may struggle to capture the full range of crack dynamics.


\section{Conclusions}

The main contribution of this paper is the introduction of a novel Ti-PD model designed to address the challenges in simulating anisotropic materials using peridynamics (PD). By incorporating a tensor into the micromodulus function, this model achieves stable numerical solutions for anisotropic materials through direct discretization, without encountering the zero-energy mode, thereby eliminating the need for additional correction methods. Moreover, the proposed model serves as an extension of the BB-PD model, offering a significantly reduced computational burden compared to the OSB-PD and NOSB-PD models. Additionally, it can revert to the BB-PD model when simulating isotropic materials with a constant Poisson's ratio.

Various examples, including models based on 2D and 3D cases, are introduced to verify the algorithm's convergence and computational efficiency. These tests confirm that the proposed model produces more stable numerical solutions with shorter computation times than the NOSB-PD model. Additionally, the model’s applicability to crack propagation was tested using discontinuous problems, demonstrating that it provides better simulation performance compared to the OSB-PD model.
\section*{Appendix A. Derivation process of $\mathcal{D}$ in 3D}
The equation in 3D model can be expressed as follows:
\begin{align}
    \dfrac{D^{1\hspace{0.12em}1}_{.1.1}}{10}+\dfrac{D^{1\hspace{0.12em}2}_{.1.2}}{30}+\dfrac{D^{1\hspace{0.12em}3}_{.1.3}}{30}=Q_{11}\\    \dfrac{D^{1\hspace{0.12em}1}_{.1.1}}{30}+\dfrac{D^{1\hspace{0.12em}2}_{.1.2}}{10}+\dfrac{D^{1\hspace{0.12em}3}_{.1.3}}{30}=Q_{44}\\
    \dfrac{D^{1\hspace{0.12em}1}_{.1.1}}{30}+\dfrac{D^{1\hspace{0.12em}2}_{.1.2}}{30}+\dfrac{D^{1\hspace{0.12em}3}_{.1.3}}{10}=Q_{55}\\
    \dfrac{D^{1\hspace{0.12em}1}_{.2.1}}{10}+\dfrac{D^{1\hspace{0.12em}2}_{.2.2}}{30}+\dfrac{D^{1\hspace{0.12em}3}_{.2.3}}{30}=Q_{14}\\
    \dfrac{D^{1\hspace{0.12em}1}_{.2.1}}{30}+\dfrac{D^{1\hspace{0.12em}2}_{.2.2}}{10}+\dfrac{D^{1\hspace{0.12em}3}_{.2.3}}{30}=Q_{24}\\
    \dfrac{D^{1\hspace{0.12em}1}_{.2.1}}{30}+\dfrac{D^{1\hspace{0.12em}2}_{.2.2}}{30}+\dfrac{D^{1\hspace{0.12em}3}_{.2.3}}{10}=Q_{56}\\
    \dfrac{D^{1\hspace{0.12em}1}_{.3.1}}{10}+\dfrac{D^{1\hspace{0.12em}2}_{.3.2}}{30}+\dfrac{D^{1\hspace{0.12em}3}_{.3.3}}{30}=Q_{15}\\
    \dfrac{D^{1\hspace{0.12em}1}_{.3.1}}{30}+\dfrac{D^{1\hspace{0.12em}2}_{.3.2}}{100}+\dfrac{D^{1\hspace{0.12em}3}_{.3.3}}{30}=Q_{46}\\
    \dfrac{D^{1\hspace{0.12em}1}_{.3.1}}{30}+\dfrac{D^{1\hspace{0.12em}2}_{.3.2}}{30}+\dfrac{D^{1\hspace{0.12em}3}_{.3.3}}{10}=Q_{35}\\
    \dfrac{2D^{1\hspace{0.12em}2}_{.1.1}}{15}=2Q_{14}\\
    \dfrac{2D^{1\hspace{0.12em}3}_{.1.1}}{15}=2Q_{15}\\
    \dfrac{2D^{1\hspace{0.12em}3}_{.1.2}}{15}=2Q_{45}\\
    \dfrac{2D^{1\hspace{0.12em}2}_{.2.1}}{15}=Q_{12}+Q_{44}\\
    \dfrac{2D^{1\hspace{0.12em}3}_{.2.1}}{15}=Q_{16}+Q_{45}\\
    \dfrac{2D^{1\hspace{0.12em}3}_{.2.2}}{15}=Q_{46}+Q_{25}\\
    \dfrac{2D^{1\hspace{0.12em}2}_{.3.1}}{15}=Q_{16}+Q_{45}\\
    \dfrac{2D^{1\hspace{0.12em}3}_{.3.1}}{15}=Q_{13}+Q_{55}\\
    \dfrac{2D^{1\hspace{0.12em}3}_{.3.2}}{15}=Q_{34}+Q_{56}
\end{align}
\begin{align}
\dfrac{D^{2\hspace{0.12em}1}_{.2.1}}{10}+\dfrac{D^{2\hspace{0.12em}2}_{.2.2}}{30}+\dfrac{D^{2\hspace{0.12em}3}_{.2.3}}{30}=Q_{44}\\
\dfrac{D^{2\hspace{0.12em}1}_{.2.1}}{30}+\dfrac{D^{2\hspace{0.12em}2}_{.2.2}}{10}+\dfrac{D^{2\hspace{0.12em}3}_{.2.3}}{30}=Q{22}\\
\dfrac{D^{2\hspace{0.12em}1}_{.2.1}}{30}+\dfrac{D^{2\hspace{0.12em}2}_{.2.2}}{30}+\dfrac{D^{2\hspace{0.12em}3}_{.2.3}}{10}=Q_{66}\\
\dfrac{D^{2\hspace{0.12em}1}_{.3.1}}{10}+\dfrac{D^{2\hspace{0.12em}2}_{.3.2}}{30}+\dfrac{D^{2\hspace{0.12em}3}_{.3.3}}{30}=Q_{45}\\
\dfrac{D^{2\hspace{0.12em}1}_{.3.1}}{30}+\dfrac{D^{2\hspace{0.12em}2}_{.3.2}}{10}+\dfrac{D^{2\hspace{0.12em}3}_{.3.3}}{30}=Q_{26}\\
\dfrac{D^{2\hspace{0.12em}1}_{.3.1}}{30}+\dfrac{D^{2\hspace{0.12em}2}_{.3.2}}{30}+\dfrac{D^{2\hspace{0.12em}3}_{.3.3}}{10}=Q_{36}\\
\dfrac{2D^{2\hspace{0.12em}2}_{.2.1}}{15}=2Q_{24}\\
\dfrac{2D^{2\hspace{0.12em}3}_{.2.1}}{15}=2Q_{46}\\
\dfrac{2D^{2\hspace{0.12em}3}_{.2.2}}{15}=2Q_{26}\\
\dfrac{2D^{2\hspace{0.12em}2}_{.3.1}}{15}=Q_{46}+Q_{25}\\
\dfrac{2D^{2\hspace{0.12em}3}_{.3.1}}{15}=Q_{34}+Q_{56}\\
\dfrac{2D^{2\hspace{0.12em}3}_{.3.2}}{15}=Q_{34}+Q_{66}\\
\dfrac{D^{3\hspace{0.12em}1}_{.3.1}}{10}+\dfrac{D^{3\hspace{0.12em}2}_{.3.2}}{30}+\dfrac{D^{3\hspace{0.12em}3}_{.3.3}}{30}=Q_{55}\\
\dfrac{D^{3\hspace{0.12em}1}_{.3.1}}{30}+\dfrac{D^{3\hspace{0.12em}2}_{.3.2}}{10}+\dfrac{D^{2\hspace{0.12em}3}_{.3.3}}{30}=Q_{66}\\
\dfrac{D^{3\hspace{0.12em}1}_{.3.1}}{30}+\dfrac{D^{3\hspace{0.12em}2}_{.3.2}}{30}+\dfrac{D^{3\hspace{0.12em}3}_{.3.3}}{10}=Q_{33}\\
\dfrac{2D^{3\hspace{0.12em}2}_{.3.1}}{15}=2Q_{56}\\
\dfrac{2D^{3\hspace{0.12em}3}_{.3.1}}{15}=2Q_{35}\\
dfrac{2D^{3\hspace{0.12em}3}_{.3.2}}{15}=2Q_{36}
\end{align}

\section*{Appendix B. Spatial discretization for 2D Ti-PD}
 The entries of stiff matrix for 2D Ti-PD can be expressed as:

     \begin{align}
& A_{2p-1, 2q-1}= \begin{cases}\dfrac{\left(3\left(3 Q_{11}-Q_{66}\right) \Delta x_{p, q}^2+3\left(3 Q_{66}-Q_{11}\right) \Delta y_{p, q}^2+24Q_{16}\Delta x_{p, q}\Delta y_{p, q}\right) V_q}{\pi \delta^3 h\left(\sqrt{\Delta x_{p, q}^2+\Delta y_{p, q}^2}\right)^3}  &, p \neq q \\ 
-\sum_{p \neq q} A_{2p-1, 2q-1} &, p=q\end{cases} \notag\\
& A_{2p-1, 2q}\hspace{0.75em} =\begin{cases}\dfrac{\left(3\left(Q_{11}-3Q_{26}\right) \Delta x_{p, q}^2+12\left(Q_{12}+Q_{66}\right) \Delta x_{p, q}\Delta y_{p, q}\right) V_q}{\pi \delta^3 h\left(\sqrt{\Delta x_{p, q}^2+\Delta y_{p, q}^2}\right)^3}  \\
+\dfrac{3\left(Q_{26}-3Q_{16}\right) \Delta y_{p, q}^2}{\pi \delta^3 h\left(\sqrt{\Delta x_{p, q}^2+\Delta y_{p, q}^2}\right)^3}&\hspace{6em}, p \neq q\\
-\sum_{p \neq q} A_{2p-1, 2q} &\hspace{6em}, p=q\end{cases} \notag\\
& A_{2p,2q}\hspace{1em}=\hspace{0.5em}  \begin{cases}\dfrac{\left(3\left(3Q_{66}-Q_{26}\right) \Delta x_{p, q}^2+3\left(3Q_{22}-Q_{66}\right) \Delta y_{p, q}^2+24Q_{26}\Delta x_{p, q}\Delta y_{p, q}\right)V_q}{\pi \delta^3 h\left(\sqrt{\Delta x_{p, q}^2+\Delta y_{p, q}^2}\right)^3} &, p \neq q \\
-\sum_{p \neq q} A_{2p, 2q} &, p=q\end{cases} \notag
\end{align}
and $A_{2p, 2q-1}=A_{2p-1, 2q}$,
where $\Delta x_{p,q}=x_q-x_p$, $\Delta y_{p,q}=y_q-y_p$, .
\section*{Appendix C. Spatial discretization for 3D Ti-PD}
 With $\Delta x_{p,q}=x_q-x_p$, $\Delta y_{p,q}=y_q-y_p$, $\Delta z_{p,q}=z_q-z_p$ The stiff matrix entries in 3D Ti-PD can be expressed as:
	\begin{align}
	&A_{3p-2,3q-2}=\begin{cases}\dfrac{(3(4Q_{11}-Q_{44}-D_{55})\Delta x_{p,q}^2+3(4Q_{44}-Q_{11}-Q_{55})\Delta y_{p,q}^2)V_q}{\pi \delta^4 (\sqrt{\Delta x_{p,q}^2+\Delta y_{p,q}^2+\Delta z^2_{p,q}})^3}\\
 +\dfrac{(3(4Q_{55}-Q_{11}-Q_{44})\Delta z_{p,q}^2+30Q_{14}\Delta x\Delta y+30Q_{15}\Delta x\Delta z)V_q}{\pi \delta^4 (\sqrt{\Delta x^2_{p,q}+\Delta y_{p,q}^2+\Delta z^2_{p,q}})^3} \vspace{0.5em} \\ 
+\dfrac{30Q_{45}\Delta y\Delta zV_q}{\pi \delta^4 (\sqrt{\Delta x^2_{p,q}+\Delta y_{p,q}^2+\Delta z^2_{p,q}})^3}&\hspace{-2em},\text{$p \neq q$}
 \\
	-\sum_{p \neq q}A_{3p-2,3q-2}&\hspace{-2em},\text{$p=q$}
	\end{cases}\notag
	\\&A_{3p-2,3q}=\begin{cases}\dfrac{(3(4Q_{15}-Q_{46}-Q_{35})\Delta x_{p,q}^2+3(4Q_{46}-Q_{15}-Q_{35})\Delta y_{p,q}^2)V_q}{\pi \delta^4 (\sqrt{\Delta x_{p,q}^2+\Delta y_{p,q}^2+\Delta z^2_{p,q}})^3}\\
 +\dfrac{(3(4Q_{35}-Q_{15}-Q_{46})+15(Q_{16}+Q_{45})\Delta x\Delta y+15(Q_{13}+Q_{55})\Delta x\Delta z)V_q}{\pi \delta^4 (\sqrt{\Delta x^2_{p,q}+\Delta y_{p,q}^2+\Delta z^2_{p,q}})^3}\\
+\dfrac{(15(Q_{34}+Q_{56})\Delta y\Delta z)V_q}{\pi \delta^4 (\sqrt{\Delta x^2_{p,q}+\Delta y_{p,q}^2+\Delta z^2_{p,q}})^3}&\hspace{-2em},\text{$p \neq q$}
 \\
	-\sum_{p \neq q}A_{3p-2,3q}&\hspace{-2em},\text{$p=q$}
	\end{cases}\notag
	\\&A_{3p-2,3q-1}=\begin{cases}\dfrac{(3(4Q_{14}-Q_{24}-Q_{56})\Delta x_{p,q}^2+3(4Q_{24}-Q_{14}-Q_{56})\Delta y_{p,q}^2)V_q}{\pi \delta^4 (\sqrt{\Delta x_{p,q}^2+\Delta y_{p,q}^2+\Delta z^2_{p,q}})^3}\\
 +\dfrac{(3(4Q_{56}-Q_{24}-Q_{14})+15(Q_{12}+Q_{44})\Delta x\Delta y+15(Q_{16}+Q_{45})\Delta x\Delta z)V_q}{\pi \delta^4 (\sqrt{\Delta x^2_{p,q}+\Delta y_{p,q}^2+\Delta z^2_{p,q}})^3}\\
+\dfrac{(15(Q_{46}+Q_{25})\Delta y\Delta z)V_q}{\pi \delta^4 (\sqrt{\Delta x^2_{p,q}+\Delta y_{p,q}^2+\Delta z^2_{p,q}})^3}&\hspace{-2em},\text{$p \neq q$}
 \\
	-\sum_{p \neq q}A_{3p-2,3q-1}&\hspace{-2em},\text{$p=q$}
	\end{cases}\notag
	\\&A_{3p-1,3q-1}=\begin{cases}\dfrac{(3(4Q_{44}-Q_{22}-Q_{66})\Delta x_{p,q}^2+3(4Q_{22}-Q_{44}-Q_{66})\Delta y_{p,q}^2)V_q}{\pi \delta^4 (\sqrt{\Delta x_{p,q}^2+\Delta y_{p,q}^2+\Delta z^2_{p,q}})^3}\\
 +\dfrac{(3(4Q_{66}-Q_{22}-Q_{44})+30Q_{24}\Delta x\Delta y+30Q_{46}\Delta x\Delta z)V_q}{\pi \delta^4 (\sqrt{\Delta x^2_{p,q}+\Delta y_{p,q}^2+\Delta z^2_{p,q}})^3}\\
+\dfrac{30Q_{26}\Delta y\Delta zV_q}{\pi \delta^4 (\sqrt{\Delta x^2_{p,q}+\Delta y_{p,q}^2+\Delta z^2_{p,q}})^3}&\hspace{-2em},\text{$p \neq q$}
 \\
	-\sum_{p \neq q}A_{3p-1,3q-1}&\hspace{-2em},\text{$p=q$}
	\end{cases}\notag\\
 &A_{3p-1,3q}=\begin{cases}\dfrac{(3(4Q_{45}-Q_{26}-Q_{36})\Delta x_{p,q}^2+3(4Q_{26}-Q_{45}-Q_{36})\Delta y_{p,q}^2)V_q}{\pi \delta^4 (\sqrt{\Delta x_{p,q}^2+\Delta y_{p,q}^2+\Delta z^2_{p,q}})^3}\\
 +\dfrac{(3(4Q_{36}-Q_{45}-Q_{26})+15(Q_{46}+Q_{25})\Delta x\Delta y+15(Q_{34}+Q_{56})\Delta x\Delta z)V_q}{\pi \delta^4 (\sqrt{\Delta x^2_{p,q}+\Delta y_{p,q}^2+\Delta z^2_{p,q}})^3}\\
+\dfrac{15(Q_{23}+Q_{66})\Delta y\Delta zV_q}{\pi \delta^4 (\sqrt{\Delta x^2_{p,q}+\Delta y_{p,q}^2+\Delta z^2_{p,q}})^3}&\hspace{-2em},\text{$p \neq q$}
 \\
	-\sum_{p \neq q}A_{3p-1,3q}&\hspace{-2em},\text{$p=q$}
	\end{cases}\notag\\
 \end{align}
 \begin{align}
 &A_{3p,3q}=\begin{cases}\dfrac{(3(4Q_{55}-Q_{66}-Q_{33})\Delta x_{p,q}^2+3(4Q_{66}-Q_{55}-Q_{33})\Delta y_{p,q}^2)V_q}{\pi \delta^4 (\sqrt{\Delta x_{p,q}^2+\Delta y_{p,q}^2+\Delta z^2_{p,q}})^3}\\
 +\dfrac{(3(4Q_{33}-Q_{55}-Q_{66})+30Q_{56}\Delta x\Delta y+30Q_{35}\Delta x\Delta z)V_q}{\pi \delta^4 (\sqrt{\Delta x^2_{p,q}+\Delta y_{p,q}^2+\Delta z^2_{p,q}})^3}\\
+\dfrac{30Q_{36}\Delta y\Delta zV_q}{\pi \delta^4 (\sqrt{\Delta x^2_{p,q}+\Delta y_{p,q}^2+\Delta z^2_{p,q}})^3}&\hspace{-2em},\text{$p \neq q$}
 \\
	-\sum_{p \neq q}A_{3p-1,3q}&\hspace{-2em},\text{$p=q$}
	\end{cases}
		\end{align}
and $A_{3p-1,3q-2}=A_{3p-2,3q-1}$, $A_{3p-1,3q}=A_{3p,3q-1}$ and $A_{3p-2,3q}=A_{3p,3q-2}$.
\section*{Declaration}

\subsubsection*{Code and Data Availability}
The custom codes generated during the current study are available from the corresponding author upon reasonable request.
No data sets were generated or analyzed during the current study.

\subsubsection*{Conflict of interest}
We declare that we have no financial or personal relationships with others or organizations that can inappropriately influence our work. No professional or other personal interest in any product, service, and/or company could be construed as influencing the position presented in, or the review of, the manuscript entitled.

   \section*{Acknowledgments}
   The first author (Hao Tian) was supported by the Fundamental Research Funds for the Central Universities (Nos. 202042008 and 202264006) and the National Natural Science Foundation of China (No. 11801533).

\bibliography{refer}
\end{document}